\documentclass[longauth]{aa}    
\usepackage[utf8]{inputenc}

\usepackage{graphicx}
\usepackage{txfonts}
\usepackage{natbib} 

\newcommand{\mjup}{M$_{\text{J}}$}

\newcommand{\rjup}{R$_{\text{J}}$}
\newcommand{\msun}{M$_{\odot}$}

\begin{document}
\title{Discovery of a brown dwarf companion to the star HIP 64892 \thanks{Based on observations collected at the European Organisation for Astronomical Research in the Southern Hemisphere under ESO programmes 096.C-0241 and 198.C-0209 (PI: J.-L. Beuzit), 098.A-9007(A) (PI: P. Sarkis). and 087.C-0790(A) (PI: M. Ireland)}} 

\author{
    A. Cheetham\inst{\ref{inst:Geneva}}
            \and
            M. Bonnefoy\inst{\ref{inst:IPAG}}
            \and
            S. Desidera\inst{\ref{inst:INAF}}
            \and
            M. Langlois\inst{\ref{inst:Lyon},\ref{inst:LAM}}
            \and
            A. Vigan\inst{\ref{inst:LAM}}
            \and
            T. Schmidt\inst{\ref{inst:LESIA}}
            \and
            J. Olofsson\inst{\ref{inst:MPIA},\ref{inst:UMIFCA},\ref{inst:Valparaiso}}
            \and
            G. Chauvin\inst{\ref{inst:IPAG},\ref{inst:UMIFCA}}
            \and
            H. Klahr\inst{\ref{inst:MPIA}}
            \and
            R. Gratton\inst{\ref{inst:INAF}}
            \and
            V. D'Orazi\inst{\ref{inst:INAF}}
            \and
            T. Henning\inst{\ref{inst:MPIA}}
            \and
            M. Janson\inst{\ref{inst:MPIA},\ref{inst:Stockholm}}
            \and
            B. Biller\inst{\ref{inst:MPIA},\ref{inst:Edinburgh}}
            \and
            S. Peretti\inst{\ref{inst:Geneva}}
            \and
            J. Hagelberg\inst{\ref{inst:Geneva},\ref{inst:IPAG}}
            \and
            D. S{\'e}gransan\inst{\ref{inst:Geneva}}
            \and
            S. Udry\inst{\ref{inst:Geneva}}
            \and
            D. Mesa\inst{\ref{inst:INAF},\ref{inst:INCT}}
            \and
            E. Sissa\inst{\ref{inst:INAF}}
            \and
            Q. Kral\inst{\ref{inst:LESIA},\ref{inst:Cambridge}}
            \and
            J. Schlieder\inst{\ref{inst:MPIA},\ref{inst:GSFC}}
            \and
            A.-L. Maire\inst{\ref{inst:MPIA}}
            \and
            C. Mordasini\inst{\ref{inst:MPIA},\ref{inst:Bern}}
            \and
            F. Menard\inst{\ref{inst:IPAG}}
            \and
            A. Zurlo\inst{\ref{inst:LAM},\ref{inst:UDP_chile}}
        \and 
            J.-L. Beuzit\inst{\ref{inst:IPAG}}
           \and 
            M. Feldt\inst{\ref{inst:MPIA}}
           \and 
            D. Mouillet\inst{\ref{inst:IPAG}}
        \and    
             M. Meyer\inst{\ref{inst:ETHZ},\ref{inst:Michigan}}
             \and
             A.-M. Lagrange\inst{\ref{inst:IPAG}}
        \and  
            A. Boccaletti\inst{\ref{inst:LESIA}}
            \and
            M. Keppler\inst{\ref{inst:MPIA}}
            \and
            T. Kopytova\inst{\ref{inst:MPIA},\ref{inst:Arizona},\ref{inst:UFU}}
            \and
            R. Ligi\inst{\ref{inst:LAM}}
            \and
            D. Rouan\inst{\ref{inst:LESIA}}
        \and 
            H. Le Coroller\inst{\ref{inst:LAM}}
            \and
            C. Dominik\inst{\ref{inst:APIA}}
            \and
            E. Lagadec\inst{\ref{inst:lagrange_nice}}
            \and
            M. Turatto\inst{\ref{inst:INAF}}
        \and         
            L. Abe\inst{\ref{inst:lagrange_nice}}
             \and 
            J. Antichi\inst{\ref{inst:INAF_firenze}}
             \and 
            A. Baruffolo\inst{\ref{inst:INAF}}
             \and 
            P. Baudoz\inst{\ref{inst:LESIA}}
             \and 
            P. Blanchard\inst{\ref{inst:LAM}}
             \and 
            T. Buey\inst{\ref{inst:LESIA}}
             \and 
            M. Carbillet\inst{\ref{inst:lagrange_nice}}
             \and 
            M. Carle\inst{\ref{inst:LAM}}
             \and 
            E. Cascone\inst{\ref{inst:INAF_napoli}}
             \and 
            R. Claudi\inst{\ref{inst:INAF}}
             \and 
            A. Costille\inst{\ref{inst:LAM}}
             \and 
            A. Delboulb\'e\inst{\ref{inst:IPAG}}
             \and 
            V. De Caprio\inst{\ref{inst:INAF_napoli}}
             \and 
            K. Dohlen\inst{\ref{inst:LAM}}
             \and 
            D. Fantinel\inst{\ref{inst:INAF}}
             \and 
            P. Feautrier\inst{\ref{inst:IPAG}}
             \and 
            T. Fusco\inst{\ref{inst:onera}}
             \and 
            E. Giro\inst{\ref{inst:INAF}}
             \and 
            L. Gluck\inst{\ref{inst:IPAG}}
             \and 
            N. Hubin\inst{\ref{inst:eso_garching}}
             \and 
            E. Hugot\inst{\ref{inst:LAM}}
             \and 
            M. Jaquet\inst{\ref{inst:LAM}}
             \and 
            M. Kasper\inst{\ref{inst:eso_garching}}
             \and 
            M. Llored\inst{\ref{inst:LAM}}
             \and 
            F. Madec\inst{\ref{inst:LAM}}
             \and 
            Y. Magnard\inst{\ref{inst:IPAG}}
             \and 
            P. Martinez\inst{\ref{inst:lagrange_nice}}
             \and 
            D. Maurel\inst{\ref{inst:IPAG}}
             \and 
            D. Le Mignant\inst{\ref{inst:LAM}}
             \and 
            O. M{\"o}ller-Nilsson\inst{\ref{inst:MPIA}}
             \and 
            T. Moulin\inst{\ref{inst:IPAG}}
             \and 
            A. Orign\'e\inst{\ref{inst:LAM}}
             \and 
            A. Pavlov\inst{\ref{inst:MPIA}}
             \and 
            D. Perret\inst{\ref{inst:LESIA}}
             \and 
            C. Petit\inst{\ref{inst:onera}}
             \and 
            J. Pragt\inst{\ref{inst:nova}}
             \and 
            P. Puget\inst{\ref{inst:IPAG}}
             \and 
            P. Rabou\inst{\ref{inst:IPAG}}
             \and 
            J. Ramos\inst{\ref{inst:MPIA}}
             \and 
            F. Rigal\inst{\ref{inst:APIA}}
             \and 
            S. Rochat\inst{\ref{inst:IPAG}}
             \and 
            R. Roelfsema\inst{\ref{inst:nova}}
             \and 
            G. Rousset\inst{\ref{inst:LESIA}}
             \and 
            A. Roux\inst{\ref{inst:IPAG}}
             \and 
            B. Salasnich\inst{\ref{inst:INAF}}
             \and 
            J.-F. Sauvage\inst{\ref{inst:onera}}
             \and 
            A. Sevin\inst{\ref{inst:LESIA}}
             \and 
            C. Soenke\inst{\ref{inst:eso_garching}}
             \and 
            E. Stadler\inst{\ref{inst:IPAG}}
             \and 
             M. Suarez\inst{\ref{inst:eso_chile}}
             \and 
            L. Weber\inst{\ref{inst:Geneva}}
             \and 
            F. Wildi\inst{\ref{inst:Geneva}}}
   \institute{D{\'e}partement d'Astronomie, Universit{\'e} de Gen{\`e}ve, 51 chemin des Maillettes, 1290, Versoix, Switzerland \email{anthony.cheetham@unige.ch} \label{inst:Geneva}
   \and 
   Universit{\'e} Grenoble Alpes, CNRS, IPAG, 38000 Grenoble, France\label{inst:IPAG}
    \and
    INAF - Osservatorio Astronomico di Padova, Vicolo dell’ Osservatorio 5, 35122, Padova, Italy\label{inst:INAF}
    \and
    CRAL, UMR 5574, CNRS, Universit{\'e} de Lyon, Ecole Normale Sup{\'e}rieure de Lyon, 46 Alle d’Italie, F-69364 Lyon Cedex 07, France\label{inst:Lyon}
    \and
    Aix Marseille Universit{\'e}, CNRS, LAM (Laboratoire d’Astrophysique de Marseille) UMR 7326, 13388 Marseille, France\label{inst:LAM}
    \and
    LESIA, Observatoire de Paris, PSL Research University, CNRS, Sorbonne Universités, UPMC Univ. Paris 06, Univ. Paris Diderot, Sorbonne Paris Cité, 5 place Jules Janssen, 92195 Meudon, France\label{inst:LESIA}
    \and
    Max Planck Institute for Astronomy, K{\"o}nigstuhl 17, D-69117 Heidelberg, Germany\label{inst:MPIA}
    \and
    Unidad Mixta Internacional Franco-Chilena de Astronom{\'i}a,CNRS/INSU UMI 3386 and Departamento de Astronom{\'i}a, Universidad de Chile, Casilla 36-D, Santiago, Chile\label{inst:UMIFCA}
    \and
    N\'ucleo Milenio Formaci\'on Planetaria - NPF, Universidad de Valpara\'iso, Av. Gran Breta\~na 1111, Valpara\'iso, Chile\label{inst:Valparaiso}
    \and
    Department of Astronomy, Stockholm University, AlbaNova University Center, SE-10691, Stockholm, Sweden\label{inst:Stockholm}
    \and
    Institute for Astronomy, University of Edinburgh, Blackford Hill View, Edinburgh EH9 3HJ, UK\label{inst:Edinburgh}
    \and
    INCT, Universidad De Atacama, calle Copayapu 485, Copiap\'{o}, Atacama, Chile\label{inst:INCT}
    \and
    Institute of Astronomy, University of Cambridge, Madingley Road,Cambridge CB3 0HA, UK\label{inst:Cambridge}
    \and
    Exoplanets and Stellar Astrophysics Laboratory, Code 667, NASA Goddard Space Flight Center, Greenbelt MD, USA\label{inst:GSFC}
    \and
    Physikalisches Institut, University of Bern, Sidlerstrasse 5, 3012 Bern, Switzerland\label{inst:Bern}
    \and
    N\'ucleo de Astronom\'ia, Facultad de Ingenier\'ia y Ciencias, Universidad Diego Portales, Av. Ejercito 441, Santiago, Chile\label{inst:UDP_chile}
    \and
    Institute for Particle Physics and Astrophysics, ETH Zurich, Wolfgang-Pauli-Strasse 27, 8093 Zurich, Switzerland\label{inst:ETHZ}
    \and
    The University of Michigan, Ann Arbor, MI 48109, USA\label{inst:Michigan}
    \and
    School of Earth \& Space Exploration, Arizona State University, Tempe AZ 85287, USA\label{inst:Arizona}
    \and
    Ural Federal University, Yekaterinburg 620002, Russia\label{inst:UFU}
    \and
    Anton Pannekoek Institute for Astronomy, Science Park 904, NL-1098 XH Amsterdam, The Netherlands\label{inst:APIA}
    \and
    Universite Cote d’Azur, OCA, CNRS, Lagrange, France\label{inst:lagrange_nice}
    \and
    INAF - Osservatorio Astrofisico di Arcetri, Largo E. Fermi 5, I-50125 Firenze, Italy\label{inst:INAF_firenze}
    \and
    INAF - Osservatorio Astronomico di Capodimonte, Salita Moiariello 16, 80131 Napoli, Italy\label{inst:INAF_napoli}
    \and
    ONERA (Office National d’Etudes et de Recherches Aérospatiales), B.P.72, F-92322 Chatillon, France\label{inst:onera}
    \and
    European Southern Observatory (ESO), Karl-Schwarzschild-Str. 2, 85748 Garching, Germany\label{inst:eso_garching}
    \and
    NOVA Optical Infrared Instrumentation Group, Oude Hoogeveensedijk 4, 7991 PD Dwingeloo, The Netherlands\label{inst:nova}
    \and
    European Southern Observatory (ESO), Alonso de Córdova 3107, Vitacura, Casilla 19001, Santiago, Chile\label{inst:eso_chile}}

\abstract{We report the discovery of a bright, brown dwarf companion to the star HIP 64892, imaged with VLT/SPHERE during the SHINE exoplanet survey. The host is a B9.5V member of the Lower-Centaurus-Crux subgroup of the Scorpius Centaurus OB association. The measured angular separation of the companion ($1.2705\pm0.0023$") corresponds to a projected distance of $159\pm12$\,AU. We observed the target with the dual-band imaging and long-slit spectroscopy modes of the IRDIS imager to obtain its SED and astrometry. In addition, we reprocessed archival NACO L-band data, from which we also recover the companion. Its SED is consistent with a young (<30\,Myr), low surface gravity object with a spectral type of M9$_{\gamma}\pm1$. From comparison with the BT-Settl atmospheric models we estimate an effective temperature of $T_{\textrm{eff}}=2600 \pm 100$\,K, and comparison of the companion photometry to the COND evolutionary models yields a mass of $\sim29-37$\,\mjup\, at the estimated age of $16^{+15}_{-7}$\,Myr for the system. HIP 64892 is a rare example of an extreme-mass ratio system ($q\sim0.01$) and will be useful for testing models relating to the formation and evolution of such low-mass objects.}


\maketitle

\section{Introduction}
While evidence suggests that the frequency of short period stellar and planetary-mass companions to main sequence stars is high, there appears to be a relative lack of companions in the brown dwarf regime \citep{2006ApJ...640.1051G}. These objects appear to exist in an overlap region of formation processes, as the low-mass tail of stellar binary formation, and as the high-mass end of the planetary distribution. Observations of brown dwarf companions to young stars then present an important opportunity to study these different formation pathways.

The processes that form companions of all masses appear to have a sensitive dependence on the host star mass, with evidence suggesting that high and intermediate mass stars have more of such companions than their lower mass counterparts \citep[e.g.][]{2010ApJ...709..396B,2010PASP..122..905J,2013ApJ...773..170J,2014A&A...566A.113J,2015ApJS..216....7B,2016A&A...596A..83L}.

However, our knowledge of such companions is limited by the small number of objects detected to date, covering a wide range of parameter space in terms of companion mass, primary mass, age and orbital semi-major axis. Only a few wide-separation objects have been detected around stars with masses $>2$\,\msun, such as HIP 78530B  \citep{2011ApJ...730...42L}, $\kappa$ And b \citep{2013ApJ...763L..32C}, HR 3549B \citep{2015ApJ...811..103M} and HIP 77900B \citep{2013ApJ...773...63A}.

A new generation of dedicated planet-finding instruments such as SPHERE \citep{2008SPIE.7014E..18B} and GPI \citep{2008SPIE.7015E..18M} offer significant improvements in the detection capability for substellar companions, as well as for spectroscopic and astrometric follow-up.

The SHINE survey \citep{2017A&A...605L...9C} utilizes the SPHERE instrument at the VLT to search the close environments of 600 young, nearby stars for substellar and planetary companions. In this paper, we present the imaging discovery and follow-up spectroscopy of a young, low mass brown dwarf companion identified during the course of this survey.

\section{Stellar Properties}
HIP 64892 is a B9.5 star \citep{houk1993}, classified as a member of the Lower Centaurus Crux (hereafter LCC) association by \cite{dezeeuw1999} and \cite{rizzuto2011}, with membership probabilities of 99\% and 74\%, respectively. A recent update to the BANYAN tool by \citet{2018arXiv180109051G} gives a membership probability of 64\% to the LCC subgroup, 33\% to the younger Upper Centaurus Lupus subgroup, and a 3\% probability of HIP 64892 being a field star. Since the star is not included in GAIA DR1, we adopt the trigonometric parallax by \citet{vl2007}, yielding a distance of $125 \pm 9$ pc. Photometry of the system is collected in Table ~\ref{tab:stellar_params}. No significant variability is reported by Hipparcos (photometric scatter 0.007 mag).

To further characterize the system, the star was observed with the FEROS spectrograph at the 2.2m MPG telescope operated by the Max Planck Institute for Astronomy. Data were obtained on 2017-03-30 and were reduced with the CERES package \citep{ceres}\footnote{https://github.com/rabrahm/ceres}. A summary of the stellar properties derived from this analysis, and those compiled from the literature, is given in Table ~\ref{tab:stellar_params}. From the FEROS spectrum, we measured a Radial Velocity (RV) of 14.9 km/s and a projected rotational velocity of 178 km/s, using the cross-correlation function (CCF) procedure described in \citet{2017A&A...605L...9C}.  This latter value is not unusual among stars of similar spectral type, unless significant projection effects are at work.

The FEROS CCF does not show any indication of additional components. This agrees with the conclusions of \citet{chini2012}, who found no evidence of binarity from five RV measurements. The observed RV values from that study were not published, preventing an assessment of possible long-term RV variability\footnote{The RV provided by SIMBAD originates from the study of \citet{madsen2002} and was not derived from spectroscopic measurements but rather from the velocity expected from kinematics.}.
The SPHERE data also do not provide any indication of the presence of bright stellar companions, ruling out an equal-luminosity binary down to a separation of about 40 mas.

By comparing the observed colours of HIP 64892 with those expected from the tables by \citet{pecaut2013}, we find they are consistent with the B9.5 spectral classification, and find a low reddening value of E(B-V)=0.01 consistent with the lack of V band extinction found by \citet{2012ApJ...756..133C}. In addition, the \citet{2012ApJ...756..133C} analysis showed no signs of an infrared excess. We explore the implications of this on the presence of dust in the system in Appendix \ref{appendix:dust_mass}.

\begin{figure}
\includegraphics[width=0.98\columnwidth]{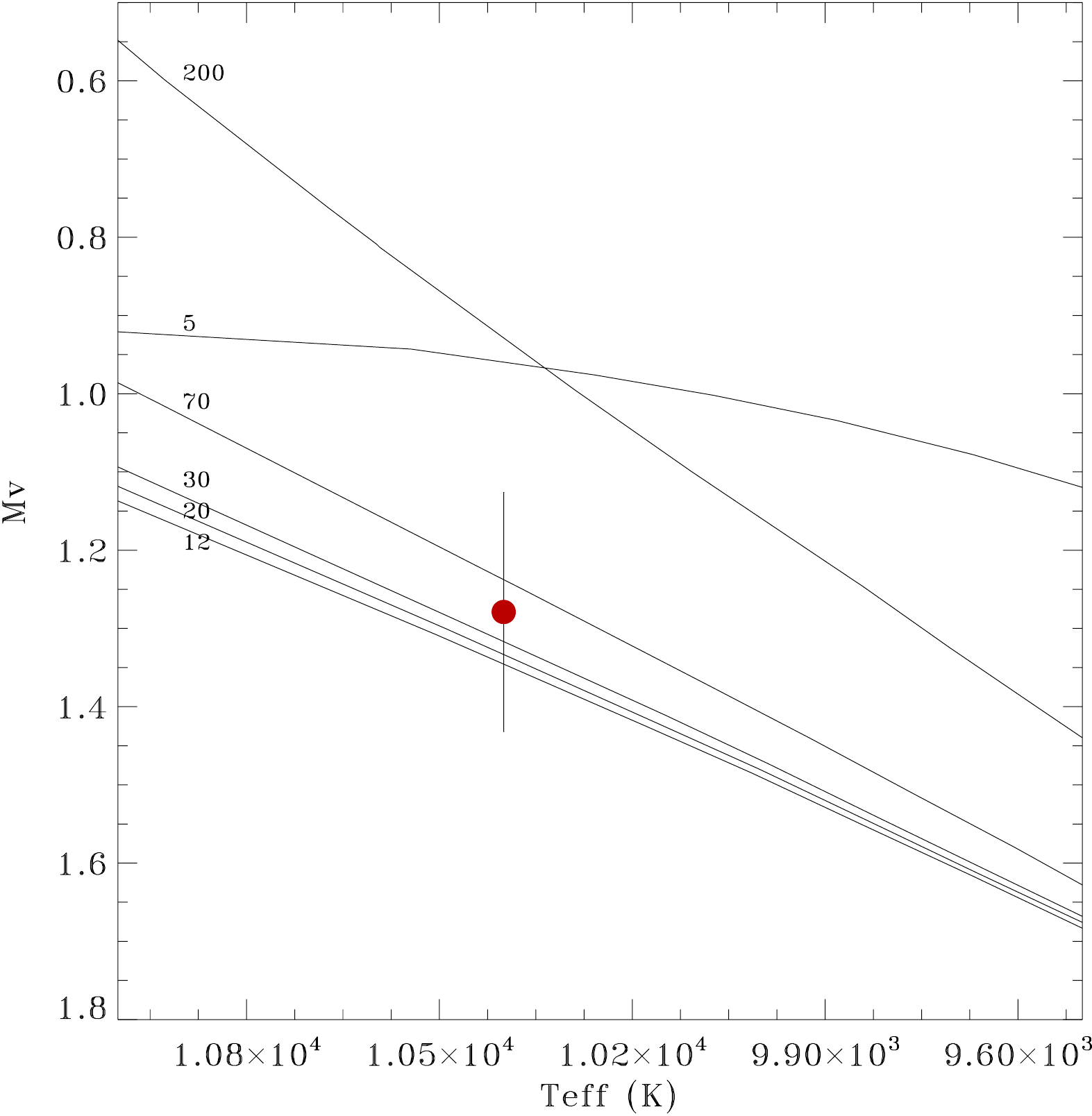}
\caption{The V-band absolute magnitude and effective temperature of HIP 64892 compared to predictions from the 5, 12, 20, 30, 70 and 200Myr isochrones of \citet{bressan2012}. The placement of HIP 64892 is consistent with the estimated local age of 16\,Myr within the measured uncertainties. }\label{fig:isochrones}
\end{figure}

The \citet{pecaut2013} tables predict an effective temperature of 10400\,K for a B9.5 star. For such a hot star, the pre-main sequence isochrones collapse on the zero-age main sequence (ZAMS) in less than 10\,Myr, and significant post-main sequence evolution is not expected for tens of Myr following this. When combined with the large uncertainty on the parallax of HIP 64892, this makes prediction of its age based on isochronal analysis difficult. The V magnitude and effective temperature of HIP 64892 are shown in Figure \ref{fig:isochrones}, relative to the \citet{bressan2012} isochrones at various ages between 5-200\,Myr. Within the uncertainties, the placement of HIP 64892 is close to the ZAMS, and therefore consistent with LCC membership.

The Sco-Cen sub-groups are known to show significant age spread. The recent age map by \citet{pecaut2016} yields a value of 16 Myr at the location of our target, equal to the commonly adopted age for the group. We adopt this value as the most likely age of HIP 64892, with the approximate ZAMS time as a lower limit. While an upper limit from comparison with the isochrones would be $\sim$100\,Myr, we instead use the nearby star TYC 7780-1467-1 to place a more precise bound on the age. This star has a well-determined age of 20\,Myr calculated from comparison with isochrones and supported by its lithium abundance \citep[EW Li=360 m\AA,][]{sacy} and fast rotation 
\citep[rotation period 4.66d,][]{kiraga2012}. Given this value, we think it unlikely that HIP 64892 is older than $\sim$30\,Myr. We adopt an uncertainty of 15\,Myr, leading to an age estimate of $16^{+15}_{-7}$.

The stellar mass and radius from the \citet{bressan2012} models are 2.35$\pm$0.09 $M_{\odot}$ and 1.79$\pm$0.10 $R_{\odot}$, respectively.

Due to its unknown rotational inclination and large measured $v\sin i$, HIP 64892 is expected to show significant oblateness. For this reason, any measurements of the stellar parameters may be influenced by its orientation. While a similar degeneracy between inclination and age caused issues for studies of other rapidly rotating stars such as $\kappa$ And \citep[e.g.][]{2013ApJ...779..153H,2016ApJ...822L...3J}, our age estimate relies on the firm Sco-Cen membership of HIP 64892, which is independent of these concerns. However, the effective temperature and mass in particular may be affected by the inclination of HIP 64892.

\begin{center}
\begin{table}
\caption{Stellar parameters of HIP 64892.}\label{tab:stellar_params}
\begin{tabular}{lcl}
\hline\hline
Parameter      & Value  & Ref \\
\hline
V (mag)                   &    6.799$\pm$0.008   & \citet{slawson1992} \\
U$-$B (mag)               &   -0.094$\pm$0.011   & \citet{slawson1992} \\
B$-$V (mag)               &   -0.018$\pm$0.006   & \citet{slawson1992} \\
V$-$I (mag)               &    0.00         & Hipparcos \\
J (mag)                   &     6.809$\pm$0.023  & \citep{cutri20032mass} \\
H (mag)                   &     6.879$\pm$0.034  & \citep{cutri20032mass} \\
K (mag)                   &     6.832$\pm$0.018  & \citep{cutri20032mass} \\
E(B-V)                    &    0.01$^{+0.02}_{-0.01}$ & this paper \\
Parallax (mas)            &    7.98$\pm$0.55 & \citet{vl2007} \\
Distance (pc)             &    $125\pm9$\,pc & from parallax\\
$\mu_{\alpha}$ (mas\,yr$^{-1}$)  & -30.83$\pm$0.50  & \citet{vl2007} \\
$\mu_{\delta}$ (mas\,yr$^{-1}$)  & -20.22$\pm$0.43  & \citet{vl2007} \\
RV   (km\,s$^{-1}$)            &  14.9  & this paper \\
SpT                            & B9.5V & \citet{houk1993}\\
$T_{\rm eff}$ (K)      &  10400 & SpT+Pecaut calib. \\
$v \sin i $  (km\,s$^{-1}$)          &   178 & this paper \\
Age (Myr)            &   $16^{+15}_{-7}$ & this paper \\
$M_{\text{star}} (M_{\odot})$   &  2.35$\pm$0.09  & this paper \\ 
$R_{\text{star}} (R_{\odot})$   &  1.79$\pm$0.10 $R_{\odot}$, & this paper \\
\hline\hline
\end{tabular}
\end{table}
\end{center}

\section{Observations and Data Reduction}
\subsection{SPHERE Imaging}
HIP 64892 was observed with SPHERE on 2016-04-01 as part of the SHINE exoplanet survey. These data were taken with the IRDIFS mode. After a bright companion candidate was discovered in these data, follow-up observations were taken on 2017-02-08 using the IRDIFS\_EXT mode to confirm its co-moving status and extend the spectral coverage.

These modes allow the Integral Field Spectrograph \citep[IFS;][]{2008SPIE.7014E..3EC} and Infra-Red Dual-band Imager and Spectrograph \citep[IRDIS;][]{2008SPIE.7014E..3LD} modules to be used simultaneously through the use of a dichroic. In these configurations, IFS provides a low resolution spectrum ($R\sim55$ across Y-J bands or $R\sim35$ across Y-H bands) while IRDIS operates in dual-band imaging mode \citep{2010MNRAS.407...71V}.

For each observing sequence, several calibration frames were taken at the beginning and end of the sequence. These consisted of unsaturated short exposure images to estimate the flux of the primary star and as a Point Spread Function (PSF) reference, followed by a sequence of images with a sinusoidal modulation introduced to the deformable mirror to generate satellite spots used to calculate the position of the star behind the coronagraph. The majority of the observing sequence consisted of long exposure (64\,s) coronagraphic imaging.

For the 2017 data, the satellite spots were used for the entirety of the coronagraphic imaging sequence rather than a separate set of frames at the beginning and end. This allowed us to correct for changes in flux and the star's position during the observations, at the cost of a small contrast loss at the separation of the satellite spots.

The data were reduced using the SPHERE Data Reduction and Handling (DRH) pipeline \citep{2008SPIE.7019E..39P} to perform the basic image cleaning steps (background subtraction, flat fielding, removal of bad pixels, calculation of the star's position behind the coronagraph, as well as extraction of the spectral data cubes for IFS). To process the IFS data, the DRH routines were augmented with additional routines from the SPHERE Data Center to reduce the spectral cross-talk and improve the wavelength calibration and bad pixel correction \citep{2015A&A...576A.121M}. Both IFS and IRDIS data were corrected using the astrometric calibration procedures described in \cite{2016arXiv160906681M}.

\begin{figure*}
\includegraphics[width=0.95\textwidth]{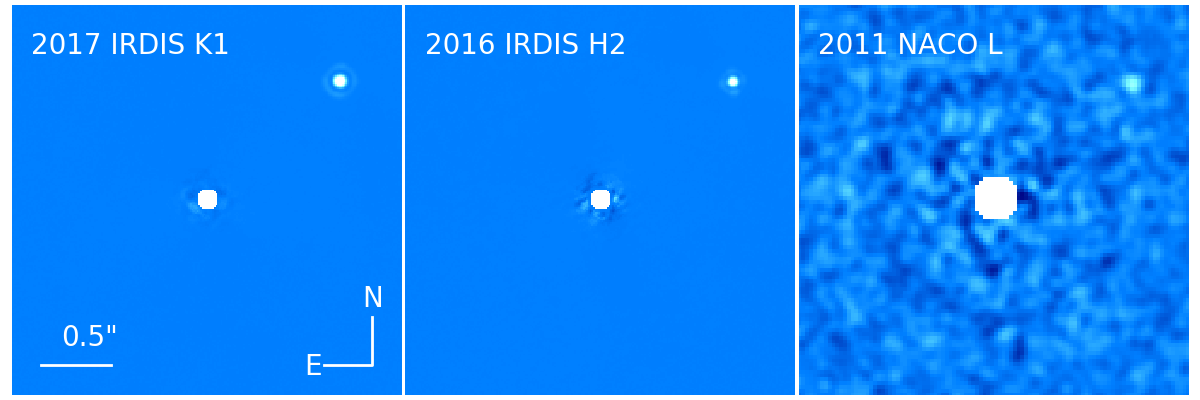}
\caption{ADI-processed images from the two SPHERE-IRDIS datasets taken with the K1 and H2 filters, and the NACO data taken with the L' filter. The companion is detected with SNR > 1000 in the two SPHERE epochs, and SNR > 8 in the NACO data.}\label{fig:images}
\end{figure*}

A bright companion candidate is clearly visible in the raw coronagraphic frames, at a separation of $1.270\pm0.002"$ in the 2016 dataset ($159\pm12$\,AU in projected physical separation). This separation places it outside of the field of view of the IFS module, and so we report the results from IRDIS only. To extract the astrometry and photometry of this object, a classical Angular Differential Imaging procedure \citep[ADI; ][]{2006ApJ...641..556M} was applied to remove the contribution from the primary star while minimising self-subtraction and other systematic effects that may be introduced by more aggressive PSF subtraction techniques. The ADI procedure and calculation of the relative astrometry and photometry of the companion were accomplished using the Specal pipeline developed for the SHINE survey (R. Galicher, 2018, in preparation). The final reduced images from this approach are shown in Fig.~\ref{fig:images}.

The astrometry was measured using the negative companion injection technique \citep{2010Sci...329...57L}, where the mean unsaturated PSF of the primary star was subtracted from the raw frames. The position and flux of the injected PSF was varied to minimise the standard deviation inside a 3 FWHM diameter region around the companion in the final ADI processed image. Once the best-fit values were found, each parameter was varied until the standard deviation increased by a factor of 1.15. This value was empirically calculated to correspond to $1\sigma$ uncertainties across a range of potential companion and observational parameters.

A systematic uncertainty of 2\,mas on the star position was adopted for the 2016 dataset, which dominates the astrometric uncertainty budget.

\begin{figure}
\includegraphics[width=0.98\columnwidth]{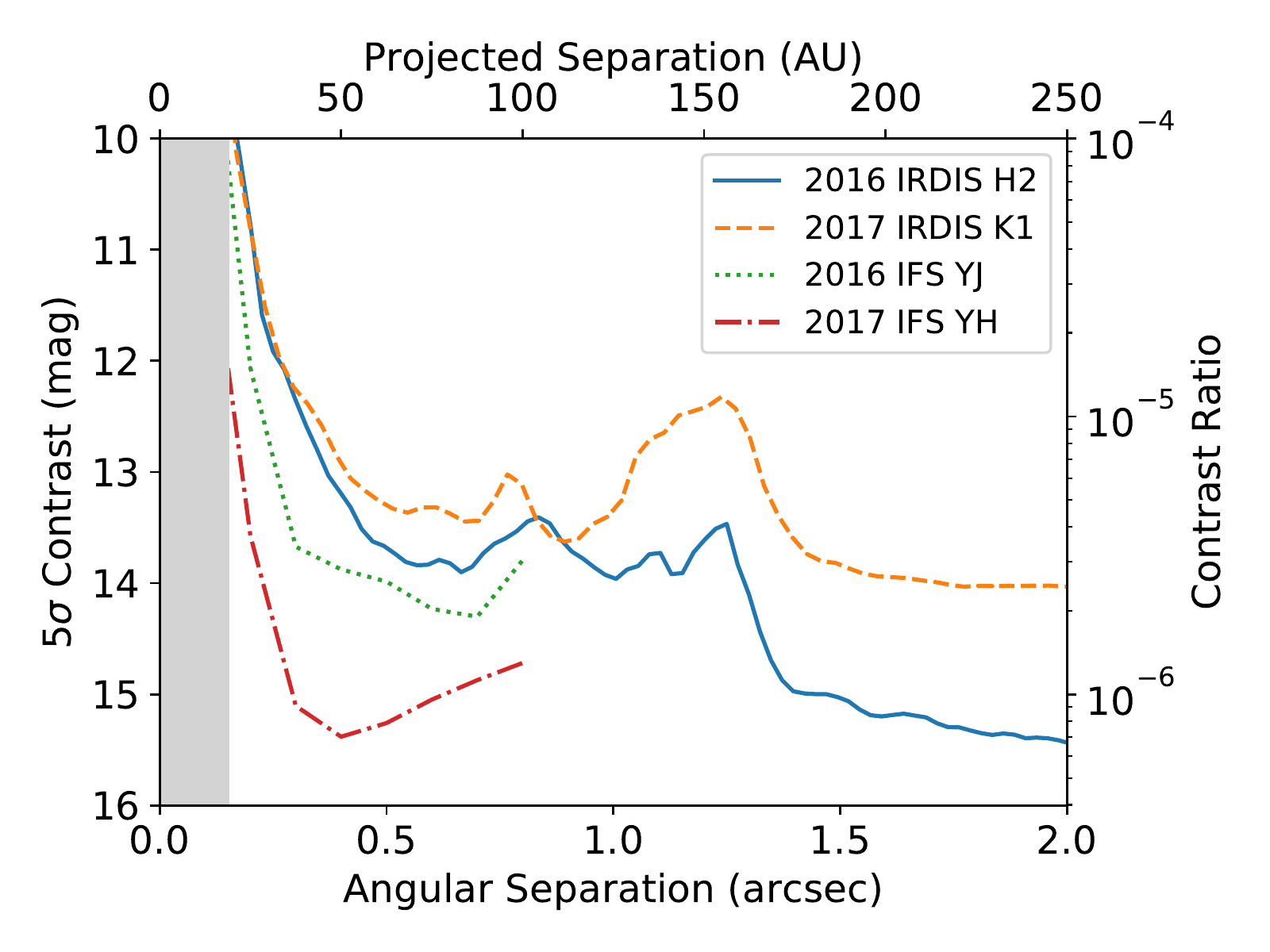}
\caption{5$\sigma$ detection limits for the two SPHERE datasets, after processing using the TLOCI algorithm. The grey shaded region indicates the separations partially or fully blocked by the coronagraph. }\label{fig:contrast_limits}
\end{figure}

A second reduction of each dataset was performed with the aim of searching for companions at higher contrasts. For the IRDIS datasets, the TLOCI algorithm was used \citep{2014IAUS..299...48M}, while for the IFS datasets we used the PCA-based ASDI algorithm described in \citet{2015A&A...576A.121M}. The resulting detection limits are are shown in Figure \ref{fig:contrast_limits}. Apart from the bright companion at $1.27"$, we find no evidence of companions at smaller separations. Three additional objects were detected at much larger separations, and are discussed in Section \ref{sec:astrometry}.

\subsection{SPHERE Long Slit Spectroscopy}
HIP 64892 was also observed with SPHERE IRDIS Long Slit Spectroscopy \citep[LSS;][]{2008A&A...489.1345V} on 2017-03-18. These observations utilised the medium resolution spectroscopy mode, which covers wavelengths from 0.95-1.65\,$\mu$m with a spectral resolution of R$\sim 350$. The observing sequence consisted of a series of alternate images with the companion inside and outside of the slit. To move the companion outside of the slit, a small offset is applied on the derotator so as to keep the star centered on the coronagraph. This strategy has been demonstrated to be very efficient to build and subtract reference images of the speckles and stellar halo while minimising the self-subtraction effects on the spectrum of the companion \citep{2016SPIE.9912E..26V}. Additional sky backgrounds were also obtained at the end of the sequence along with an unsaturated spectrum of the star to serve as reference for the contrast.

The data were reduced using the SILSS pipeline \citep{2016ascl.soft03001V}. This pipeline combines recipes from the standard ESO pipeline with custom {\it IDL} routines. Briefly, data are background subtracted, flat fielded and corrected for bad pixels. The wavelength calibration is performed and the data are corrected for the slight tilt of the grism, which causes a change in the position of the PSF with wavelength. Finally, the speckles are subtracted using principal component analysis, with the modes constructed from the spectra obtained with the companion outside of the slit.

\subsection{Archival NACO Sparse Aperture Masking Data}
In addition to the SPHERE data, we utilised archival sparse aperture masking (SAM) data from the VLT-NACO instrument, taken on 2011-06-08 (Program 087.C-0790(A), PI: Ireland). While the main purpose of the SAM mode is to detect companions and resolve structures at small angular separations (typically $<$300\,mas), this does not preclude the detection of bright objects at larger separations. The data were taken with the $L'$ filter in pupil tracking mode, and were split into blocks of 1600 exposures of 0.2s each. Each of the 5 blocks was interspersed amongst observations of other targets.

Rather than process the data in an interferometric framework, we treated it as a traditional ADI dataset to focus on larger separations. The data were processed using the GRAPHIC pipeline \citep{2016MNRAS.455.2178H}. The data were sky subtracted, flat fielded, cleaned of bad pixels, centred on the primary star, and stacked by binning 200 frames at a time. A Gaussian fit was performed to each PSF to calculate the position of the star, since it provides a reasonable match to the core of the SAM PSF. A python implementation of the KLIP algorithm \citep{2012ApJ...755L..28S} was then applied, where the first 15 modes were removed, resulting in the redetection of the SPHERE companion. The final reduced image is shown in the right panel of Fig. \ref{fig:images}.

To calculate the position and flux of the companion, we used the negative companion injection technique applied to the data cube after binning. We chose to minimise the square of the residuals within a circle of radius $\lambda/D$ centred on the companion peak calculated from the first reduced image. To explore the likelihood function, we used {\it emcee} \citep{Foreman-Mackey2013emcee}, a python implementation of the affine-invariant MCMC ensemble sampler. Due to the lack of detailed study about potential systematic biases introduced into the companion astrometry by applying this technique to SAM data, we conservatively added an additional uncertainty to the companion position of 1 pixel (27.1\,mas). Of particular concern is the way in which the time-varying fine structure of the large SAM PSF may influence the calculation of the star's position.

\begin{table*} 
\caption{Observing Log}  
\label{tab:observing_log}  
\center
\begin{tabular}{ccccccccccc}
\hline
UT Date & Instrument & Filter & DIT$^a$ & $\text{N}_{\text{exp}}^{a}$ & $\Delta \pi^{a}$ & True North correction & Plate Scale \\
& & & [s] & & [$^{\circ}$] & [$^{\circ}$] & [mas/pixel] \\
\hline
2011-06-08 & NACO  & L' & 0.2 & $8000$ & 130.4 & $-0.5\pm0.1$ & $27.1\pm0.05$ \\
2016-04-01 & IRDIS & H2/H3 & $64$ & 64 & 45 & $-1.73\pm0.06$ & $12.255\pm0.009$ \\
2016-04-01 & IFS   & YJ    & $64$ & 64 & 45 & $-102.21\pm0.06$ & $7.46\pm 0.02$ \\
2017-02-08 & IRDIS & K1/K2 & 64 & 72 & 61 & $-1.71\pm0.06$ & $12.249\pm0.009$ \\
2017-02-08 & IFS   & YH    & $64$ & 72 & 61 & $-102.19\pm0.06$ & $7.46\pm 0.02$ \\
\hline                  
\end{tabular}
\tablefoot{$^a$DIT refers to the integration time of each image, $\text{N}_{\text{exp}}$ to the total number of images obtained, $\Delta \pi$ to the parallactic angle range during the
sequence}
\end{table*}

\section{Results}
\subsection{Astrometry}\label{sec:astrometry}
\begin{figure*}
\includegraphics[width=0.95\textwidth]{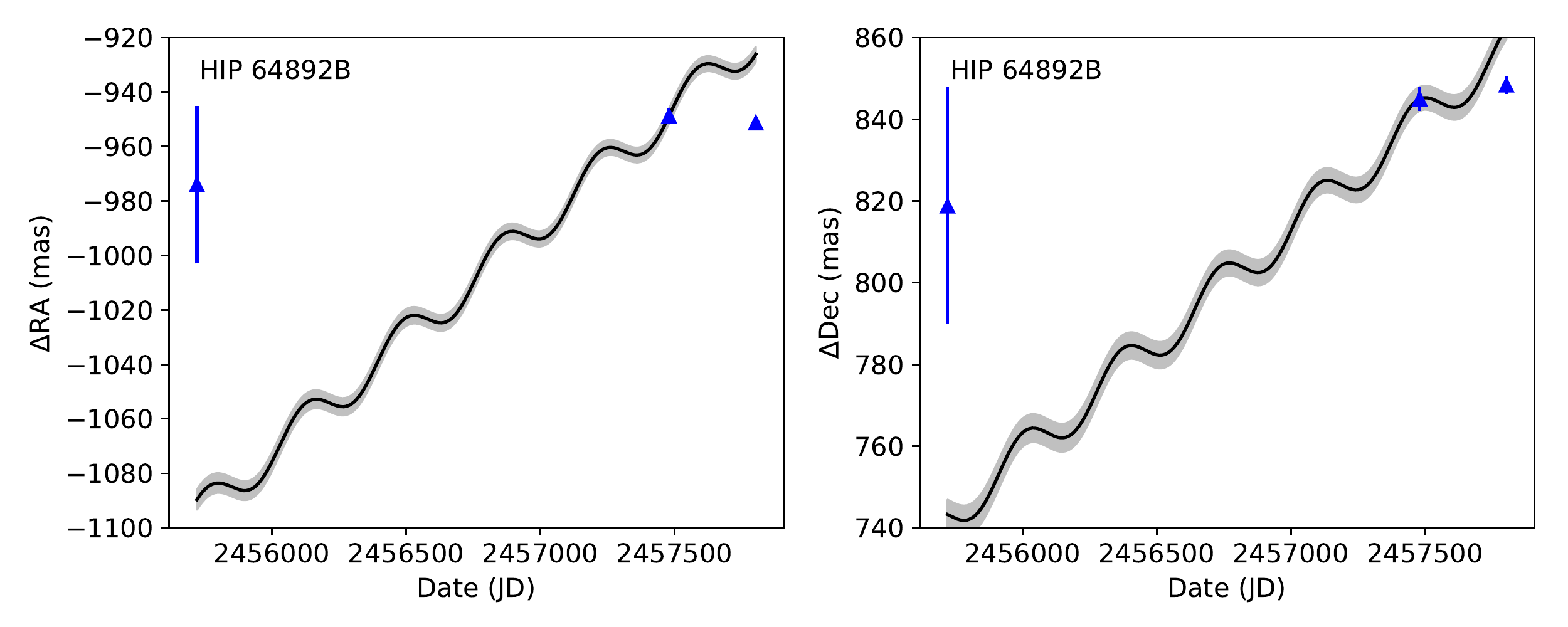}
\caption{The measured position of HIP 64892B relative to HIP 64892A. The 2011 NACO L', 2016 IRDIS H2 and 2017 IRDIS K1 positions are shown with blue triangles. The predicted motion for a stationary background object relative to the 2016 position is marked by the black line, with its uncertainty represented by the grey shaded region. The observed positions of HIP 64892B strongly conflict with the predictions for a background object, suggesting that it is co-moving. }\label{fig:proper_motion}
\end{figure*}

To confirm the co-moving status of HIP 64892B, we compared its position relative to HIP 64892A between the datasets. We used the astrometry from the 2016 SPHERE dataset with the parallax and proper motion of HIP 64892 to predict the position expected for a background object at the 2011 and 2017 epochs. The result is shown in Figure \ref{fig:proper_motion}.

For HIP 64892B, the observed positions in 2011 and 2017 differ from the prediction for a background object by 4$\sigma$ and 8$\sigma$ respectively. Instead we find a lack of significant relative motion. This clearly shows that the object is co-moving with HIP 64892A.

The angular separation of HIP 64892B corresponds to a projected separation of $159\pm12$\,AU. With the primary star mass of 2.35\,M$_{\odot}$ we would expect an orbital period of order $\sim10^3$\,yr. This is consistent with the lack of significant motion in the measured astrometry, and suggests that an additional epoch in 2018 or later may show clear orbital motion and help to constrain the orbital parameters of the companion.

\begin{figure*}
\includegraphics[width=0.95\textwidth]{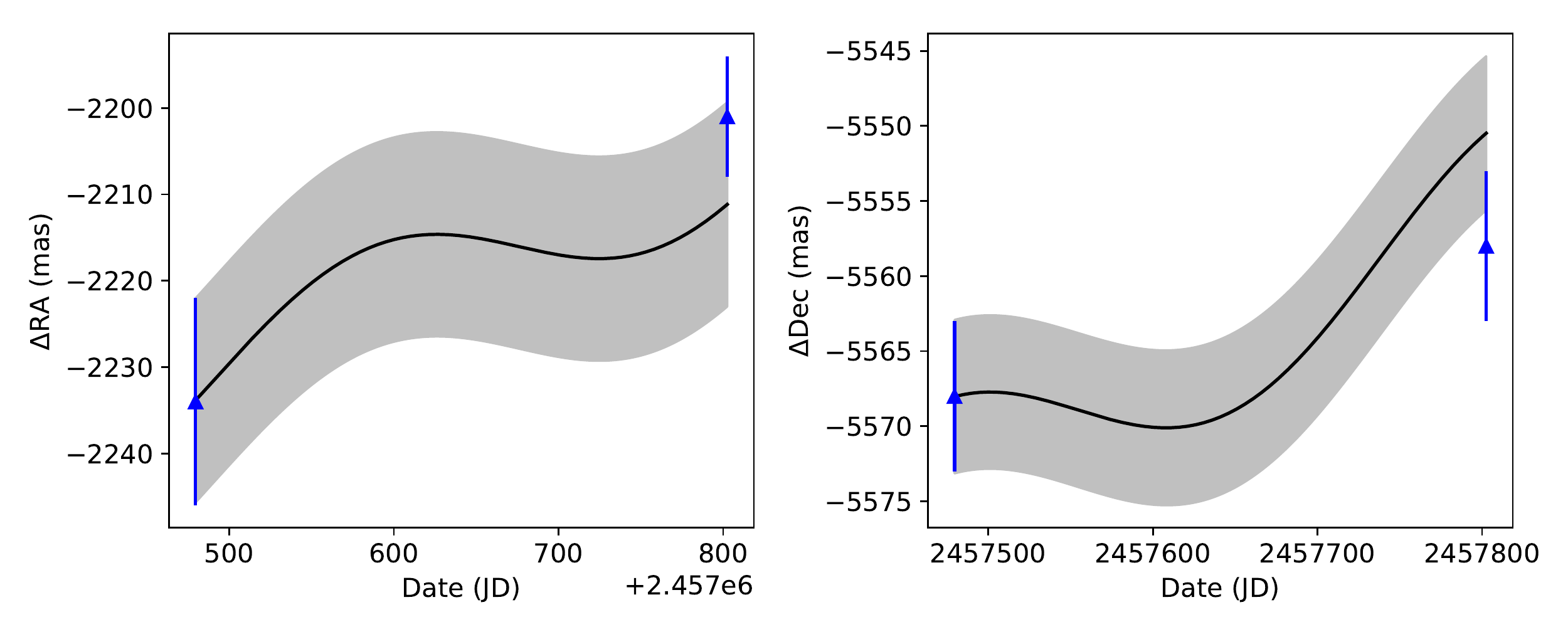}
\caption{Same as Fig. \ref{fig:proper_motion}, showing the measured position of companion candidate 1 with respect to HIP 64892A. This object agrees with the prediction for a background object.} \label{fig:proper_motion_cc1}
\end{figure*}

In addition to HIP 64892B, three objects were detected at larger separations in the IRDIS field of view. Their astrometry and photometry are given in Table \ref{tab:background_astrophotometry}. Candidate 1 was detected at high significance in both epochs, while the other two were not detected in the 2017-02-08 K band data. The relative motion of candidate 1 between the two epochs is shown in Figure \ref{fig:proper_motion_cc1} and is similar to that expected from a background star. Its position differs from the prediction by 1.3$\sigma$, while it is inconsistent with a co-moving object at 2.8$\sigma$, showing that it is likely a background star.

The IRDIS H2 and H3 photometry of the remaining two objects indicates that they are also likely background stars. If located at the same distance as HIP 64892, their H2 magnitudes and H2-H3 colours would be inconsistent with those of other known objects. Their H2 magnitudes would be consistent with those of T dwarfs, but without significant methane absorption that would be measurable in their H2-H3 colours.

\begin{table*} 
\caption{Observed Astrometry and Photometry of HIP 64892 B}
\label{tab:astrophotometry}  
\center
\begin{tabular}{cccccccc}
\hline
UT Date & Instrument & Filter & $\rho$ (") & $\theta$ (\degr) & Contrast (mag) & Abs. mag &  Mass (COND)\\
\hline
2011-06-08 & NACO  & L' & $1.272\pm0.029$   & $310.0\pm1.3$   & $6.10\pm0.08$ & $7.61\pm0.17$ & $37\pm9$ \\
2016-04-01 & IRDIS & H2 & $1.2703\pm0.0023$ & $311.68\pm0.15$ & $7.23\pm0.08$ & $8.73\pm0.17$ & $29\pm4$ \\
2016-04-01 & IRDIS & H3 & $1.2704\pm0.0022$ & $311.69\pm0.15$ & $6.99\pm0.08$ & $8.46\pm0.17$ & $29\pm5$ \\
2017-02-08 & IRDIS & K1 & $1.2753\pm0.0010$ & $311.74\pm0.12$ & $6.80\pm0.08$ & $8.29\pm0.17$ & $34\pm7$ \\
2017-02-08 & IRDIS & K2 & $1.2734\pm0.0010$ & $311.77\pm0.12$ & $6.49\pm0.12$ & $7.97\pm0.19$ & $35\pm8$ \\
\hline                  
\end{tabular}
\end{table*}

\begin{table*}
\caption{Additional objects detected by SPHERE-IRDIS}
\label{tab:background_astrophotometry}  
\center
\begin{tabular}{ccccccccc}
\hline
Candidate & UT Date &  Filter & $\rho$ (mas) & $\theta$ (deg) & $\Delta$RA (mas) & $\Delta$Dec (mas) &  Contrast (mag) \\
\hline
1 & 2016-04-01 & H2 & $6000\pm3$ & $201.86\pm0.12$ & $-2234\pm12$ & $-5568\pm5$ & $10.5\pm0.1$\\
1 & 2017-02-08 & K1 & $5978\pm4$ & $201.60\pm0.07$ & $-2201\pm7$ & $-5558\pm5$ & $10.7\pm0.1$\\
2 & 2016-04-01 & H2 & $5865\pm5$ & $202.86\pm0.12$ & $-2278\pm12$ & $-5404\pm7$ & $13.7\pm0.1$\\
3 & 2016-04-01 & H2 & $7003\pm6$ & $186.89\pm0.12$ & $-840\pm14$ & $-6952\pm6$ & $10.2\pm0.1$\\
\hline                  
\end{tabular}
\end{table*}

\subsection{Mass}
To estimate the mass of HIP 64892B, we first converted each of the apparent photometric fluxes for the companion (IRDIS H2, H3, K1, K2 and NACO L') into absolute magnitudes using the distance of $125\pm9$\,pc. To calculate the photometric zeropoint of each filter, we used the spectrum of Vega from \cite{2007ASPC..364..315B} along with the filter transmission profiles from each instrument.

We then interpolated the COND evolutionary models \citep{2003A&A...402..701B} using the age and absolute magnitudes to yield estimates of the mass of HIP 64892B. The measurements, listed in Table \ref{tab:astrophotometry}, are consistent with masses of 29-37\,\mjup. This implies a mass ratio between the brown dwarf and the primary of $q\sim0.014$.

\subsection{Spectral Properties}

\begin{figure*}
\includegraphics[width=0.95\textwidth]{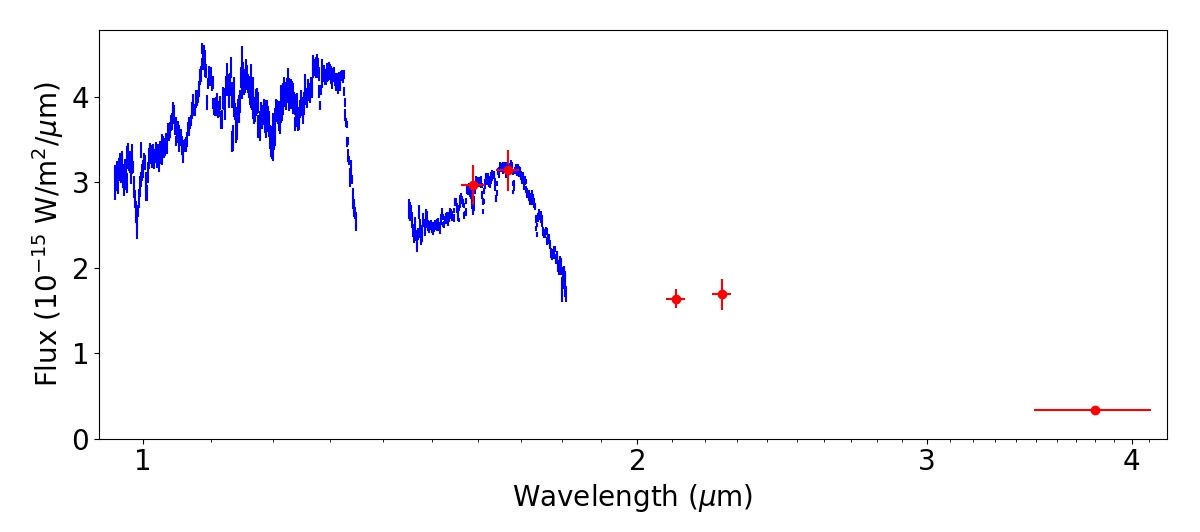}
\caption{The observed spectrum of HIP 64892B. SPHERE IRDIS LSS data are shown in blue, while the SPHERE IRDIS and NACO photometric measurements are in red. The errorbars on the x-axis of the photometric measurements represent the FWHM of the filter used.}\label{fig:companion_spectrum}
\end{figure*}

We produced a spectral model for the primary by scaling a BT-Settl spectrum with $T_{\rm eff}=10400$\,K, $\log g=4$, [M/H]$=-0.5$ using photometric measurements compiled from 2MASS, Tycho-2 and WISE \citep{skrutskie2006two,2000AA...355L..27H,2010AJ....140.1868W}. This was then used to convert the contrast measurements from IRDIS and NACO to apparent fluxes. The resulting spectrum is shown in Figure \ref{fig:companion_spectrum}.

To estimate the spectral type and effective temperature of the companion, we compared the observed spectrum and photometry of HIP 64892B with a range of spectra compiled from the literature, using the goodness-of-fit statistic $G$ \citep[e.g.][]{2008ApJ...678.1372C}. We considered young L dwarfs from the Upper-Scorpius subgroup \citep{2008MNRAS.383.1385L} as well as companions of Upper Scorpius stars \citep{2008ApJ...689L.153L,2015ApJ...802...61L}. We also compared the companion spectrum to those of   young free floating objects from the Montreal\footnote{https://jgagneastro.wordpress.com/the-montreal-spectral-library/}  \citep{2016ApJ...830..144R,2014ApJ...785L..14G,2014ApJ...792L..17G,2015ApJ...808L..20G} and \citet{2013ApJ...772...79A} libraries, and to  libraries of medium-resolution spectra of  old MLT field dwarfs \citep{2003ApJ...596..561M,2005ApJ...623.1115C,2009ApJS..185..289R,2002ApJ...564..421B}.

\begin{figure}
\includegraphics[width=0.98\columnwidth]{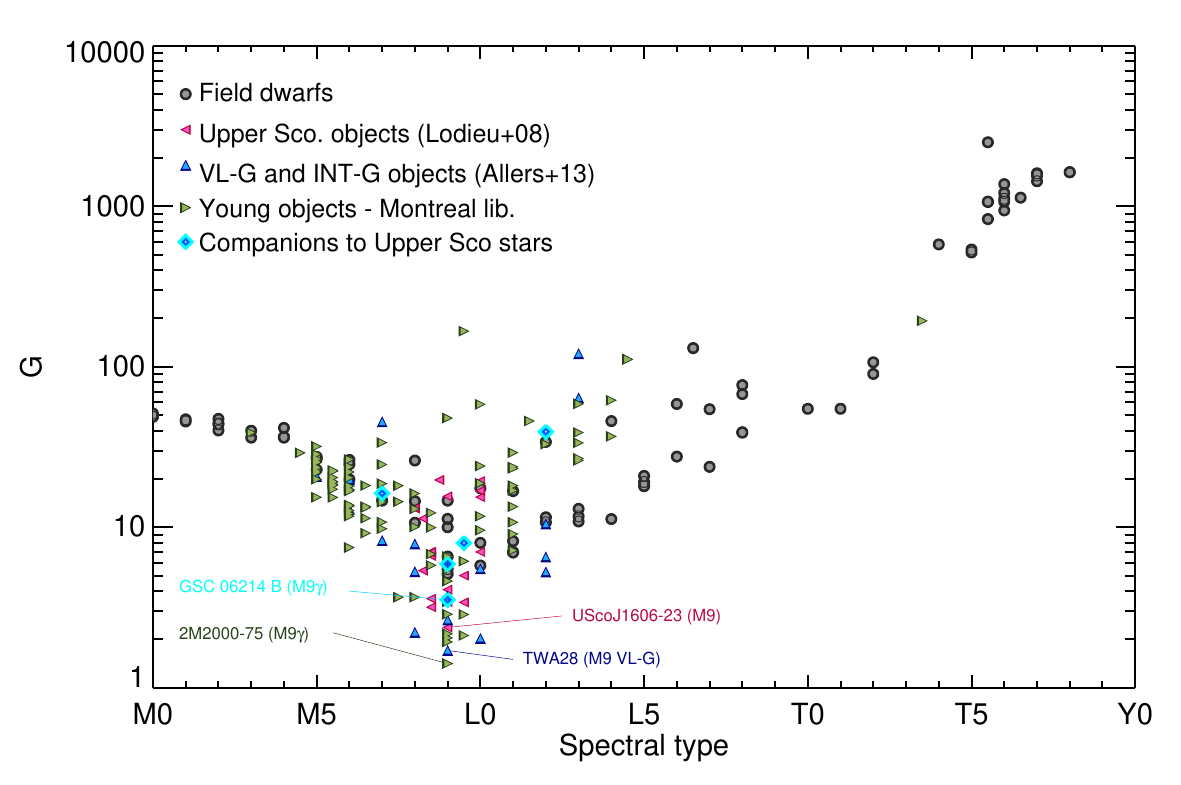}
\caption{The $G$ goodness-of-fit statistic plotted as a function of spectral type for a number of objects in the range M0-T8. A clear $G$ minimum is seen at late M spectral types and the best matches are given by young, low surface gravity objects with spectral types of M9.}\label{fig:spectral_type}
\end{figure}

The result is plotted in Fig. \ref{fig:spectral_type}, for a range of objects covering different ages, masses and spectral types. We find the best matches are given by young, low surface gravity objects with spectral types close to M9. From this, we adopt a spectral type of M9$\pm$1 for HIP 64892B.

\begin{figure}
\includegraphics[width=0.98\columnwidth]{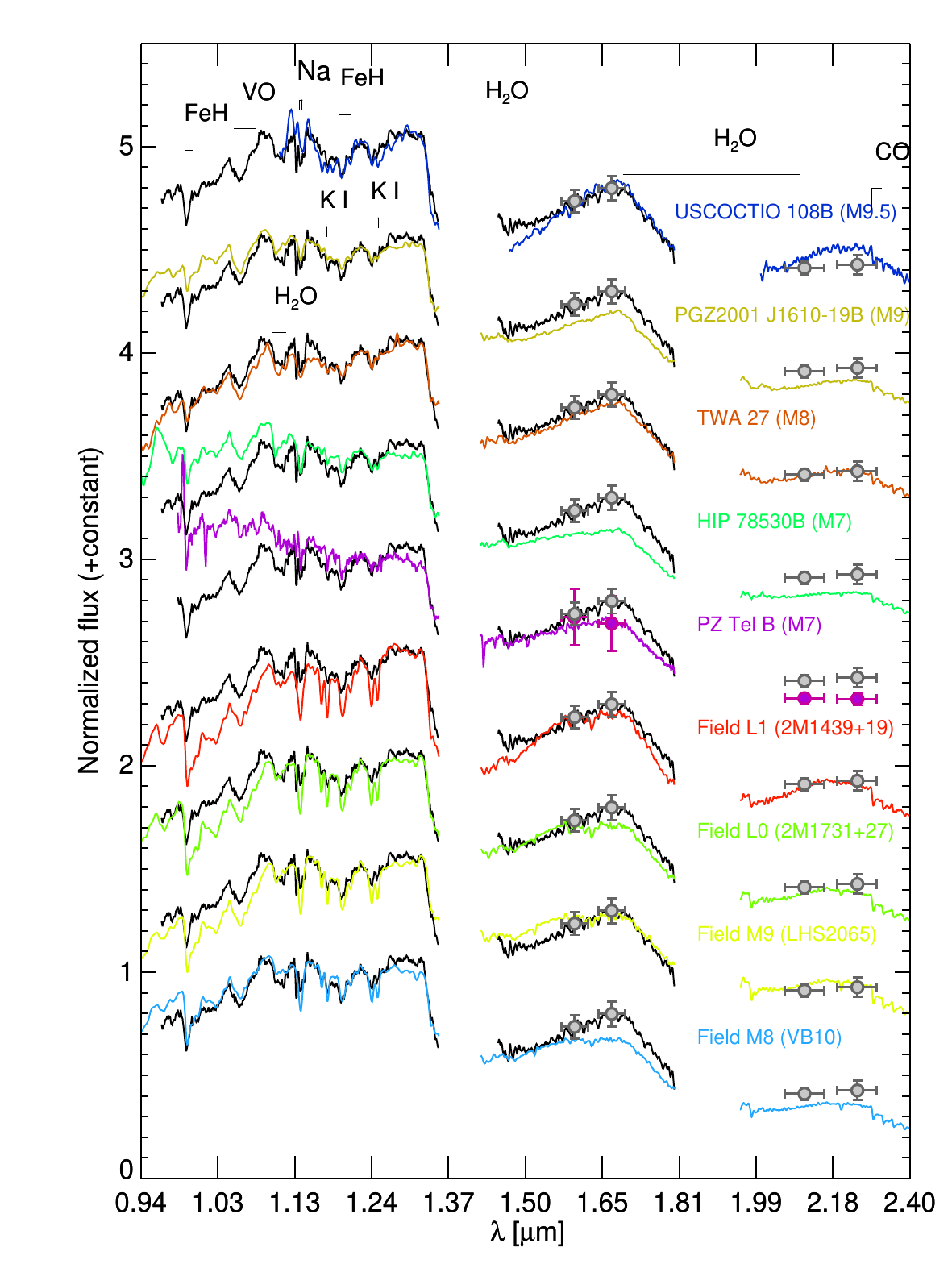}
\caption{Comparison of the observed spectrum (in black) with field dwarfs and similar young low-mass brown dwarf companions (coloured lines). The measured photometric points from the IRDIS filters are shown with grey circles. The M9 object 2MASS J20004841–7523070 (2M2000-75) provides the best match to the measured spectrum of HIP 64892B.}\label{fig:spectral_comparison}
\end{figure}

In Fig. \ref{fig:spectral_comparison}, we show the comparison of the observed spectrum of HIP 64892B to those of young and old field dwarfs at the M/L transition. Several features in the spectrum are indicative of a young object. The doublets of gravity-sensitive potassium bands at 1.169/1.177\,$\mu$m and 1.243/1.253\,$\mu$m  are reduced. The H-band has a triangular shape characteristic of low gravity atmospheres. Visually, the spectrum agrees well with those of M9$_{\gamma}$ candidate members of the $\beta$ Pictoris and Argus moving groups \citep[20-50 Myr;][]{2015ApJS..219...33G}, with 2MASS J20004841–7523070 giving the best fit.

As seen in Fig. \ref{fig:spectral_comparison}, the spectrum of HIP 64892B is also well reproduced by the spectrum of the  8 Myr-old  M8 dwarf 2M1207A (TWA27) from the TW Hydrae association, which shows many of the same features seen in our data. When compared to the SPHERE-LSS data of the M7 companion PZ Tel B \citep{2016A&A...587A..56M}, we can see a clear difference in slope indicating a later spectral type for HIP 64892B.
Also shown are the objects UScoCTIO 108B \citep{2008ApJ...673L.185B}, HIP 78530B \citep{2011ApJ...730...42L} and several field dwarfs.

Using the empirical relation between spectral type and effective temperature for young objects with low surface gravity from \cite{2015ApJ...810..158F}, our spectral type constraints correspond to an effective temperature of $T_{\text{eff}}=2600\pm300$\,K.

From the flux-calibrated LSS spectrum, we derived a synthetic absolute magnitude of $M_{\text{J,2MASS}} = 9.37 \pm 0.15$\,mag. Combining this with the spectral type of M9$\pm$1, we can use the bolometric correction relations from \cite{2015ApJ...810..158F} to derive a bolometric luminosity of $\log(L/L_\odot)= -2.66\pm0.10$ dex. In addition, we converted the K1 flux measurement to a predicted $K_{\text{S,2MASS}}$ absolute magnitude using the SpeX spectrum of TWA 27A as an analog. The resulting value of $M_{\text{Ks,2MASS}} = 8.02 \pm 0.17$\,mag corresponds to a luminosity of $\log(L/L_\odot)= -2.51\pm0.11$ dex using the same method.

We also compared the observed spectrum to the BT-Settl model grid \citep{2015A&A...577A..42B} as a function of $T_{\text{eff}}$, $\log g$ and radius $R$. We find a best fit temperature and radius of $T_{\text{eff}}=2600 \pm 100$\,K and $R=2.3 \pm 0.14$\,\rjup, similar to those predicted by the COND models for a $33$\,\mjup\, object with an age of 16\,Myr. We find that the $\log g$ value is poorly constrained, with a best fit value $\log g=5.5$. When combined with the radius, this predicts a much larger-than-expected mass. However, gravity-sensitive features are generally narrow and fitting to the entire spectrum at once may complicate this measurement. To investigate this, we performed the same fit to the two sections of the LSS spectrum either side of the $1.4\mu$m telluric feature individually. The Y-J band spectrum gives an estimate of $\log g=4.0\pm0.5$, while the H band spectrum yields $\log g=3.5\pm1.0$, indicating that a lower value is likely.

\begin{figure}
\includegraphics[width=0.98\columnwidth]{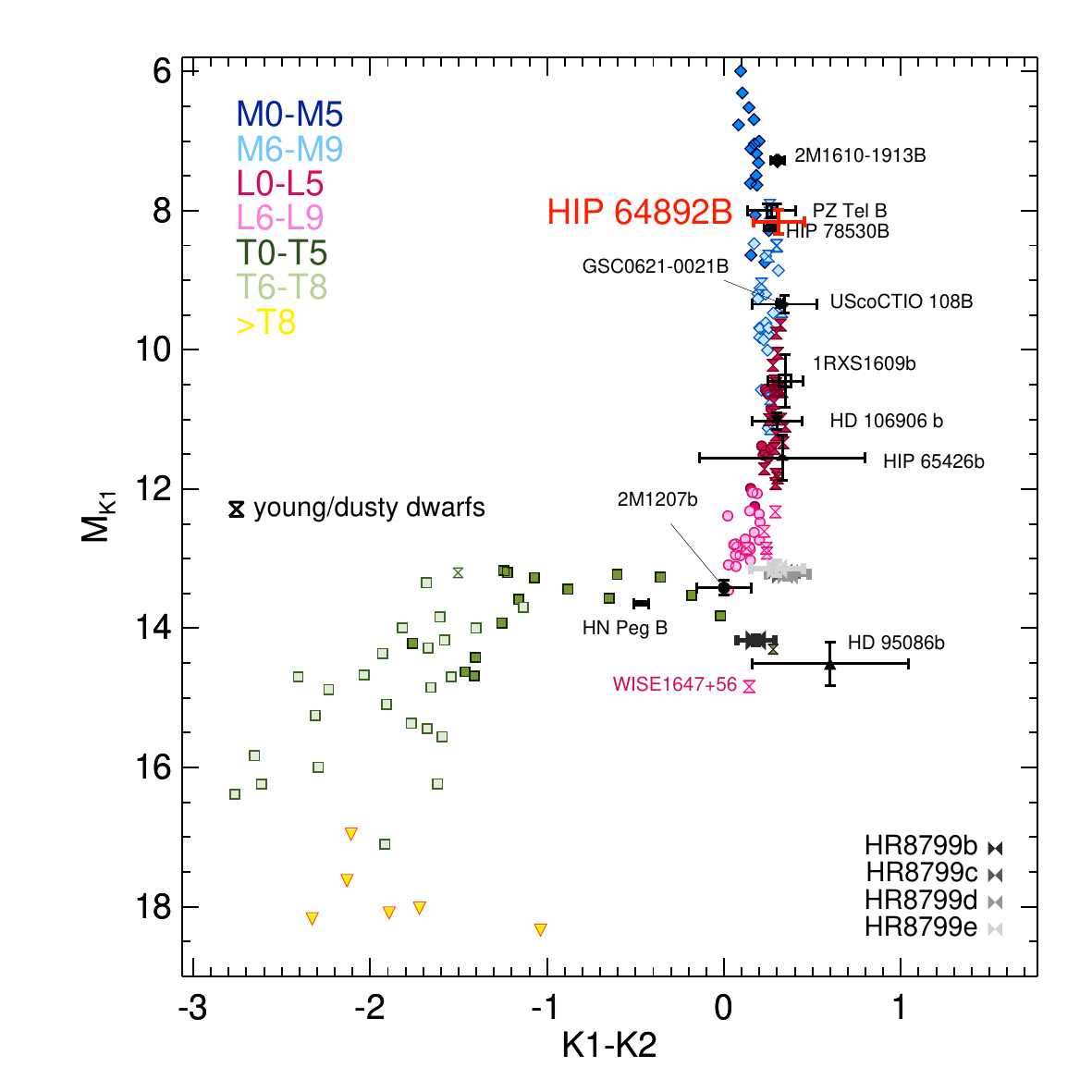}
\caption{A colour-magnitude diagram comparing the K1 absolute magnitudes and K1-K2 colours of a range of low-mass stellar and substellar objects from the literature. Synthetic K1 and K2 fluxes were computed for each target from published spectra. A number of young and dusty companions observed with SPHERE or with published K band spectra were added to the diagram. HIP 64892B falls in a similar location to PZ Tel B and HIP 78530B, both young brown dwarfs with late-M spectral types. }\label{fig:cmd_k1k2}
\end{figure}

The spectral type estimate and low surface gravity are also supported by the position of HIP 64892B on colour-magnitude diagrams. In Figure \ref{fig:cmd_k1k2} we show its K1 magnitude and K1-K2 colour compared to a range of field objects assembled from the SpeX Prism Library \citep{2014ASInC..11....7B} and from \cite{2000ApJ...535..965L}. For these objects we generated synthetic photometry using the SPHERE filter bandpasses. In addition, we also show a range of young companions with SPHERE K1K2 photometry or K-band spectra from the literature \citep{2007ApJ...654..570L,2008ApJ...689L.153L,2010A&A...517A..76P,2014A&A...562A.127B,2015ApJ...802...61L,2016A&A...586L...8L,2016A&A...587A..56M,2016A&A...587A..57Z,2017A&A...605L...9C,2018arXiv180105850C}.

The position of HIP 64892B is similar to PZ Tel B and HIP 78530B, both young companions with late-M spectral types. All three lie close to the mid-M sequence of field dwarfs, showing a slight over-luminosity compared to late-M field objects that match their spectral types. This trend is also seen in young field brown dwarfs \citep{2012ApJ...752...56F,2013AN....334...85L}.

\section{Discussion and Conclusion}
HIP 64892 joins a growing number of high or intermediate mass stars with extreme mass ratio companions at large separations ($<10\%$, $>10$\,AU). While brown dwarf companions to solar-type stars from both RV and imaging surveys are inherently rare \citep[the so-called ``brown dwarf desert'';][]{2006ApJ...640.1051G,kraus08,2009ApJS..181...62M}, evidence suggests that the occurrence rates of companions around intermediate mass stars may be substantially higher \citep{2012A&A...544A...9V}. This trend is seen for companions with masses spanning the planetary to stellar regimes \citep[e.g.][]{2010PASP..122..905J,2013ApJ...773..170J}.

Despite this, HIP 64892 is one of the highest mass stars around which a substellar companion has been detected, due to the challenges associated with observing such stars and the tendency for large surveys to focus on solar-like or low-mass stars. Only HIP 78530B \citep{2011ApJ...730...42L}, $\kappa$ And b \citep{2013ApJ...763L..32C} and HIP 77900B \citep{2013ApJ...773...63A} orbit stars with a larger mass.

The large mass of the primary leads to a mass ratio for HIP 64892B that is particularly small ($q\sim0.014$). For low-mass and solar-like stars, such a value would correspond to objects at or below the deuterium burning limit, making HIP 64892B a valuable object for studying the overlapping processes of binary star and planet formation in the brown dwarf regime.
Indeed, the properties of HIP 64892B raise the possibility that it formed via gravitational instability, either through a binary-star mechanism or through disk instability in the circumstellar disk of the primary \citep{1997Sci...276.1836B}. We investigate the latter idea further in Appendix \ref{app:gi_models}, finding that the observed mass and separation of HIP 64892B are compatible with in-situ formation via disk instability. This process is one of the proposed pathways for giant planet formation, making HIP 64892B an important object for understanding this mechanism.

HIP 64892B also stands to be an important object for calibrating substellar formation and evolutionary models, with a well-known age tied to that of the LCC association, and its brightness allowing high SNR spectroscopic and photometric measurements.

While the spectroscopic and photometric measurements presented here have high SNR, the uncertainty on the distance is relatively large, and dominates the uncertainties on the absolute magnitudes, luminosity, and isochronal mass for HIP 64892B. A much more precise distance measurement will come from the GAIA parallax, which should be included in the second data release.

Many of the derived properties of the newly discovered companion match those of the planet-hosting brown dwarf TWA 27, with an excellent match found between their spectra. This object may prove to be a useful analogue, and a point of direct comparison between young brown dwarfs in single and multiple systems.

The HIP 64892 system as a whole has many parallels to that of $\kappa$ And. While the stellar hosts have a similar mass and spectral type, the companion HIP 64892B appears to be a hotter, younger and higher-mass analogue of $\kappa$ And B \citep{2013ApJ...779..153H,2014A&A...562A.111B,2016ApJ...822L...3J}. When combined with HIP 78530B \citep{2011ApJ...730...42L}, HR 3549B \citep{2015ApJ...811..103M}, HD 1160 \citep{2012ApJ...750...53N} and $\eta$ Tel B \citep{2000ApJ...541..390L}, they form a useful sample to explore the properties of low and intermediate-mass brown dwarf companions to young, 2-3\,\msun stars.

\begin{acknowledgements}
This work has been carried out within the frame of the National Centre for Competence in Research ``PlanetS'' supported by the Swiss National Science Foundation (SNSF).\\
SPHERE is an instrument designed and built by a consortium consisting of IPAG (Grenoble, France), MPIA (Heidelberg, Germany), LAM (Marseille, France), LESIA (Paris, France), Laboratoire Lagrange (Nice, France), INAF - Osservatorio di Padova (Italy), Observatoire Astronomique de l'Université de Genève (Switzerland), ETH Zurich (Switzerland), NOVA (Netherlands), ONERA (France) and ASTRON (Netherlands) in collaboration with ESO. SPHERE was funded by ESO, with additional contributions from CNRS (France), MPIA (Germany), INAF (Italy), FINES (Switzerland) and NOVA (Netherlands). SPHERE also received funding from the European Commission Sixth and Seventh Framework Programmes as part of the Optical Infrared Coordination Network for Astronomy (OPTICON) under grant number RII3-Ct-2004-001566 for FP6 (2004-2008), grant number 226604 for FP7 (2009-2012) and grant number 312430 for FP7 (2013-2016). \\
This work has made use of the SPHERE Data Centre, jointly operated by OSUG/IPAG (Grenoble), PYTHEAS/LAM/CeSAM (Marseille), OCA/Lagrange (Nice) and Observatoire de Paris/LESIA (Paris) and supported by a grant from Labex OSUG@2020 (Investissements d’avenir – ANR10 LABX56).
This publication makes use of data products from the Two Micron All Sky Survey, which is a joint project of the University of Massachusetts and the Infrared Processing and Analysis Center/California Institute of Technology, funded by the National Aeronautics and Space Administration and the National Science Foundation.\\
This publication makes use of VOSA, developed under the Spanish Virtual Observatory project supported from the Spanish MICINN through grant AyA2011-24052.\\
R.\,G., R.\,C., S.\,D. acknowledge support from the “Progetti Premiali” funding scheme of the Italian Ministry of Education, University, and Research.\\
J.\,O. acknowledges financial support from ICM N\'ucleo Milenio de Formaci\'on Planetaria, NPF.\\
Q.\,K. acknowledges funding from STFC via the Institute of Astronomy, Cambridge Consolidated Grant.

\end{acknowledgements}

\bibliographystyle{aa.bst}

\bibliography{big_bibliography.bib}

\begin{thebibliography}{95}
\expandafter\ifx\csname natexlab\endcsname\relax\def\natexlab#1{#1}\fi

\bibitem[{{Aller} {et~al.}(2013){Aller}, {Kraus}, {Liu}, {Burgett}, {Chambers},
  {Hodapp}, {Kaiser}, {Magnier}, \& {Price}}]{2013ApJ...773...63A}
{Aller}, K.~M., {Kraus}, A.~L., {Liu}, M.~C., {et~al.} 2013, \apj, 773, 63

\bibitem[{{Allers} \& {Liu}(2013)}]{2013ApJ...772...79A}
{Allers}, K.~N. \& {Liu}, M.~C. 2013, \apj, 772, 79

\bibitem[{{Baehr} {et~al.}(2017){Baehr}, {Klahr}, \&
  {Kratter}}]{2017ApJ...848...40B}
{Baehr}, H., {Klahr}, H., \& {Kratter}, K.~M. 2017, \apj, 848, 40

\bibitem[{{Baraffe} {et~al.}(2003){Baraffe}, {Chabrier}, {Barman}, {Allard}, \&
  {Hauschildt}}]{2003A&A...402..701B}
{Baraffe}, I., {Chabrier}, G., {Barman}, T.~S., {Allard}, F., \& {Hauschildt},
  P.~H. 2003, \aap, 402, 701

\bibitem[{{Baraffe} {et~al.}(2015){Baraffe}, {Homeier}, {Allard}, \&
  {Chabrier}}]{2015A&A...577A..42B}
{Baraffe}, I., {Homeier}, D., {Allard}, F., \& {Chabrier}, G. 2015, \aap, 577,
  A42

\bibitem[{{Bayo} {et~al.}(2008){Bayo}, {Rodrigo}, {Barrado Y Navascu{\'e}s},
  {Solano}, {Guti{\'e}rrez}, {Morales-Calder{\'o}n}, \&
  {Allard}}]{2008A&A...492..277B}
{Bayo}, A., {Rodrigo}, C., {Barrado Y Navascu{\'e}s}, D., {et~al.} 2008, \aap,
  492, 277

\bibitem[{{B{\'e}jar} {et~al.}(2008){B{\'e}jar}, {Zapatero Osorio},
  {P{\'e}rez-Garrido}, {{\'A}lvarez}, {Mart{\'{\i}}n}, {Rebolo},
  {Vill{\'o}-P{\'e}rez}, \& {D{\'{\i}}az-S{\'a}nchez}}]{2008ApJ...673L.185B}
{B{\'e}jar}, V.~J.~S., {Zapatero Osorio}, M.~R., {P{\'e}rez-Garrido}, A.,
  {et~al.} 2008, \apjl, 673, L185

\bibitem[{{Beuzit} {et~al.}(2008){Beuzit}, {Feldt}, {Dohlen}, {Mouillet},
  {Puget}, {Wildi}, {Abe}, {Antichi}, {Baruffolo}, {Baudoz}, {Boccaletti},
  {Carbillet}, {Charton}, {Claudi}, {Downing}, {Fabron}, {Feautrier},
  {Fedrigo}, {Fusco}, {Gach}, {Gratton}, {Henning}, {Hubin}, {Joos}, {Kasper},
  {Langlois}, {Lenzen}, {Moutou}, {Pavlov}, {Petit}, {Pragt}, {Rabou}, {Rigal},
  {Roelfsema}, {Rousset}, {Saisse}, {Schmid}, {Stadler}, {Thalmann}, {Turatto},
  {Udry}, {Vakili}, \& {Waters}}]{2008SPIE.7014E..18B}
{Beuzit}, J.-L., {Feldt}, M., {Dohlen}, K., {et~al.} 2008, in \procspie, Vol.
  7014

\bibitem[{{Bohlin}(2007)}]{2007ASPC..364..315B}
{Bohlin}, R.~C. 2007, in Astronomical Society of the Pacific Conference Series,
  Vol. 364, The Future of Photometric, Spectrophotometric and Polarimetric
  Standardization, ed. C.~{Sterken}, 315

\bibitem[{{Bonnefoy} {et~al.}(2014{\natexlab{a}}){Bonnefoy}, {Chauvin},
  {Lagrange}, {Rojo}, {Allard}, {Pinte}, {Dumas}, \&
  {Homeier}}]{2014A&A...562A.127B}
{Bonnefoy}, M., {Chauvin}, G., {Lagrange}, A.-M., {et~al.} 2014{\natexlab{a}},
  \aap, 562, A127

\bibitem[{{Bonnefoy} {et~al.}(2014{\natexlab{b}}){Bonnefoy}, {Currie},
  {Marleau}, {Schlieder}, {Wisniewski}, {Carson}, {Covey}, {Henning}, {Biller},
  {Hinz}, {Klahr}, {Marsh Boyer}, {Zimmerman}, {Janson}, {McElwain},
  {Mordasini}, {Skemer}, {Bailey}, {Defr{\`e}re}, {Thalmann}, {Skrutskie},
  {Allard}, {Homeier}, {Tamura}, {Feldt}, {Cumming}, {Grady}, {Brandner},
  {Helling}, {Witte}, {Hauschildt}, {Kandori}, {Kuzuhara}, {Fukagawa}, {Kwon},
  {Kudo}, {Hashimoto}, {Kusakabe}, {Abe}, {Brandt}, {Egner}, {Guyon}, {Hayano},
  {Hayashi}, {Hayashi}, {Hodapp}, {Ishii}, {Iye}, {Knapp}, {Matsuo}, {Mede},
  {Miyama}, {Morino}, {Moro-Martin}, {Nishimura}, {Pyo}, {Serabyn}, {Suenaga},
  {Suto}, {Suzuki}, {Takahashi}, {Takami}, {Takato}, {Terada}, {Tomono},
  {Turner}, {Watanabe}, {Yamada}, {Takami}, \& {Usuda}}]{2014A&A...562A.111B}
{Bonnefoy}, M., {Currie}, T., {Marleau}, G.-D., {et~al.} 2014{\natexlab{b}},
  \aap, 562, A111

\bibitem[{{Boss}(1997)}]{1997Sci...276.1836B}
{Boss}, A.~P. 1997, Science, 276, 1836

\bibitem[{{Bowler} {et~al.}(2010){Bowler}, {Johnson}, {Marcy}, {Henry}, {Peek},
  {Fischer}, {Clubb}, {Liu}, {Reffert}, {Schwab}, \&
  {Lowe}}]{2010ApJ...709..396B}
{Bowler}, B.~P., {Johnson}, J.~A., {Marcy}, G.~W., {et~al.} 2010, \apj, 709,
  396

\bibitem[{{Bowler} {et~al.}(2015){Bowler}, {Liu}, {Shkolnik}, \&
  {Tamura}}]{2015ApJS..216....7B}
{Bowler}, B.~P., {Liu}, M.~C., {Shkolnik}, E.~L., \& {Tamura}, M. 2015, \apjs,
  216, 7

\bibitem[{{Brahm} {et~al.}(2017){Brahm}, {Jord{\'a}n}, \& {Espinoza}}]{ceres}
{Brahm}, R., {Jord{\'a}n}, A., \& {Espinoza}, N. 2017, \pasp, 129, 034002

\bibitem[{{Bressan} {et~al.}(2012){Bressan}, {Marigo}, {Girardi}, {Salasnich},
  {Dal Cero}, {Rubele}, \& {Nanni}}]{bressan2012}
{Bressan}, A., {Marigo}, P., {Girardi}, L., {et~al.} 2012, MNRAS, 427, 127

\bibitem[{{Burgasser}(2014)}]{2014ASInC..11....7B}
{Burgasser}, A.~J. 2014, in Astronomical Society of India Conference Series,
  Vol.~11, Astronomical Society of India Conference Series

\bibitem[{{Burgasser} {et~al.}(2002){Burgasser}, {Kirkpatrick}, {Brown},
  {Reid}, {Burrows}, {Liebert}, {Matthews}, {Gizis}, {Dahn}, {Monet}, {Cutri},
  \& {Skrutskie}}]{2002ApJ...564..421B}
{Burgasser}, A.~J., {Kirkpatrick}, J.~D., {Brown}, M.~E., {et~al.} 2002, \apj,
  564, 421

\bibitem[{{Burns} {et~al.}(1979){Burns}, {Lamy}, \&
  {Soter}}]{1979Icar...40....1B}
{Burns}, J.~A., {Lamy}, P.~L., \& {Soter}, S. 1979, \icarus, 40, 1

\bibitem[{{Carson} {et~al.}(2013){Carson}, {Thalmann}, {Janson}, {Kozakis},
  {Bonnefoy}, {Biller}, {Schlieder}, {Currie}, {McElwain}, {Goto}, {Henning},
  {Brandner}, {Feldt}, {Kandori}, {Kuzuhara}, {Stevens}, {Wong}, {Gainey},
  {Fukagawa}, {Kuwada}, {Brandt}, {Kwon}, {Abe}, {Egner}, {Grady}, {Guyon},
  {Hashimoto}, {Hayano}, {Hayashi}, {Hayashi}, {Hodapp}, {Ishii}, {Iye},
  {Knapp}, {Kudo}, {Kusakabe}, {Matsuo}, {Miyama}, {Morino}, {Moro-Martin},
  {Nishimura}, {Pyo}, {Serabyn}, {Suto}, {Suzuki}, {Takami}, {Takato},
  {Terada}, {Tomono}, {Turner}, {Watanabe}, {Wisniewski}, {Yamada}, {Takami},
  {Usuda}, \& {Tamura}}]{2013ApJ...763L..32C}
{Carson}, J., {Thalmann}, C., {Janson}, M., {et~al.} 2013, \apjl, 763, L32

\bibitem[{{Chauvin} {et~al.}(2017){Chauvin}, {Desidera}, {Lagrange}, {Vigan},
  {Gratton}, {Langlois}, {Bonnefoy}, {Beuzit}, {Feldt}, {Mouillet}, {Meyer},
  {Cheetham}, {Biller}, {Boccaletti}, {D'Orazi}, {Galicher}, {Hagelberg},
  {Maire}, {Mesa}, {Olofsson}, {Samland}, {Schmidt}, {Sissa}, {Bonavita},
  {Charnay}, {Cudel}, {Daemgen}, {Delorme}, {Janin-Potiron}, {Janson},
  {Keppler}, {Le Coroller}, {Ligi}, {Marleau}, {Messina}, {Molli{\`e}re},
  {Mordasini}, {M{\"u}ller}, {Peretti}, {Perrot}, {Rodet}, {Rouan}, {Zurlo},
  {Dominik}, {Henning}, {Menard}, {Schmid}, {Turatto}, {Udry}, {Vakili}, {Abe},
  {Antichi}, {Baruffolo}, {Baudoz}, {Baudrand}, {Blanchard}, {Bazzon}, {Buey},
  {Carbillet}, {Carle}, {Charton}, {Cascone}, {Claudi}, {Costille}, {Deboulbe},
  {De Caprio}, {Dohlen}, {Fantinel}, {Feautrier}, {Fusco}, {Gigan}, {Giro},
  {Gisler}, {Gluck}, {Hubin}, {Hugot}, {Jaquet}, {Kasper}, {Madec}, {Magnard},
  {Martinez}, {Maurel}, {Le Mignant}, {M{\"o}ller-Nilsson}, {Llored}, {Moulin},
  {Orign{\'e}}, {Pavlov}, {Perret}, {Petit}, {Pragt}, {Puget}, {Rabou},
  {Ramos}, {Rigal}, {Rochat}, {Roelfsema}, {Rousset}, {Roux}, {Salasnich},
  {Sauvage}, {Sevin}, {Soenke}, {Stadler}, {Suarez}, {Weber}, {Wildi},
  {Antoniucci}, {Augereau}, {Baudino}, {Brandner}, {Engler}, {Girard}, {Gry},
  {Kral}, {Kopytova}, {Lagadec}, {Milli}, {Moutou}, {Schlieder},
  {Szul{\'a}gyi}, {Thalmann}, \& {Wahhaj}}]{2017A&A...605L...9C}
{Chauvin}, G., {Desidera}, S., {Lagrange}, A.-M., {et~al.} 2017, \aap, 605, L9

\bibitem[{{Chauvin} {et~al.}(2018){Chauvin}, {Gratton}, {Bonnefoy}, {Lagrange},
  {de Boer}, {Vigan}, {Beust}, {Lazzoni}, {Boccaletti}, {Galicher}, {Desidera},
  {Delorme}, {Keppler}, {Lannier}, {Maire}, {Mesa}, {Meunier}, {Kral},
  {Henning}, {Menard}, {Moor}, {Avenhaus}, {Bazzon}, {Janson}, {Beuzit},
  {Bhowmik}, {Bonavita}, {Borgniet}, {Brandner}, {Cheetham}, {Cudel}, {Feldt},
  {Fontanive}, {Ginski}, {Hagelberg}, {Janin-Potiron}, {Lagadec}, {Langlois},
  {Le Coroller}, {Messina}, {Meyer}, {Mouillet}, {Peretti}, {Perrot}, {Rodet},
  {Samland}, {Sissa}, {Olofsson}, {Salter}, {Schmidt}, {Zurlo}, {Milli}, {van
  Boekel}, {Quanz}, {Wilson}, {Feautrier}, {Le Mignant}, {Perret}, {Ramos}, \&
  {Rochat}}]{2018arXiv180105850C}
{Chauvin}, G., {Gratton}, R., {Bonnefoy}, M., {et~al.} 2018, ArXiv e-prints
  [\eprint[arXiv]{1801.05850}]

\bibitem[{{Chen} {et~al.}(2012){Chen}, {Pecaut}, {Mamajek}, {Su}, \&
  {Bitner}}]{2012ApJ...756..133C}
{Chen}, C.~H., {Pecaut}, M., {Mamajek}, E.~E., {Su}, K.~Y.~L., \& {Bitner}, M.
  2012, \apj, 756, 133

\bibitem[{{Chini} {et~al.}(2012){Chini}, {Hoffmeister}, {Nasseri}, {Stahl}, \&
  {Zinnecker}}]{chini2012}
{Chini}, R., {Hoffmeister}, V.~H., {Nasseri}, A., {Stahl}, O., \& {Zinnecker},
  H. 2012, \mnras, 424, 1925

\bibitem[{{Claudi} {et~al.}(2008){Claudi}, {Turatto}, {Gratton}, {Antichi},
  {Bonavita}, {Bruno}, {Cascone}, {De Caprio}, {Desidera}, {Giro}, {Mesa},
  {Scuderi}, {Dohlen}, {Beuzit}, \& {Puget}}]{2008SPIE.7014E..3EC}
{Claudi}, R.~U., {Turatto}, M., {Gratton}, R.~G., {et~al.} 2008, in \procspie,
  Vol. 7014, Ground-based and Airborne Instrumentation for Astronomy II, 70143E

\bibitem[{{Cushing} {et~al.}(2008){Cushing}, {Marley}, {Saumon}, {Kelly},
  {Vacca}, {Rayner}, {Freedman}, {Lodders}, \& {Roellig}}]{2008ApJ...678.1372C}
{Cushing}, M.~C., {Marley}, M.~S., {Saumon}, D., {et~al.} 2008, \apj, 678, 1372

\bibitem[{{Cushing} {et~al.}(2005){Cushing}, {Rayner}, \&
  {Vacca}}]{2005ApJ...623.1115C}
{Cushing}, M.~C., {Rayner}, J.~T., \& {Vacca}, W.~D. 2005, \apj, 623, 1115

\bibitem[{{Cutri} {et~al.}(2003){Cutri}, {Skrutskie}, {van Dyk}, {Beichman},
  {Carpenter}, {Chester}, {Cambresy}, {Evans}, {Fowler}, {Gizis}, {Howard},
  {Huchra}, {Jarrett}, {Kopan}, {Kirkpatrick}, {Light}, {Marsh}, {McCallon},
  {Schneider}, {Stiening}, {Sykes}, {Weinberg}, {Wheaton}, {Wheelock}, \&
  {Zacarias}}]{cutri20032mass}
{Cutri}, R.~M., {Skrutskie}, M.~F., {van Dyk}, S., {et~al.} 2003, {2MASS All
  Sky Catalog of point sources.}

\bibitem[{{de Zeeuw} {et~al.}(1999){de Zeeuw}, {Hoogerwerf}, {de Bruijne},
  {Brown}, \& {Blaauw}}]{dezeeuw1999}
{de Zeeuw}, P.~T., {Hoogerwerf}, R., {de Bruijne}, J.~H.~J., {Brown}, A.~G.~A.,
  \& {Blaauw}, A. 1999, \aj, 117, 354

\bibitem[{{Dent} {et~al.}(2014){Dent}, {Wyatt}, {Roberge}, {Augereau},
  {Casassus}, {Corder}, {Greaves}, {de Gregorio-Monsalvo}, {Hales}, {Jackson},
  {Hughes}, {Lagrange}, {Matthews}, \& {Wilner}}]{2014Sci...343.1490D}
{Dent}, W.~R.~F., {Wyatt}, M.~C., {Roberge}, A., {et~al.} 2014, Science, 343,
  1490

\bibitem[{{Dohlen} {et~al.}(2008){Dohlen}, {Langlois}, {Saisse}, {Hill},
  {Origne}, {Jacquet}, {Fabron}, {Blanc}, {Llored}, {Carle}, {Moutou}, {Vigan},
  {Boccaletti}, {Carbillet}, {Mouillet}, \& {Beuzit}}]{2008SPIE.7014E..3LD}
{Dohlen}, K., {Langlois}, M., {Saisse}, M., {et~al.} 2008, in \procspie, Vol.
  7014, Ground-based and Airborne Instrumentation for Astronomy II, 70143L

\bibitem[{{Draine}(2003)}]{2003ARA&A..41..241D}
{Draine}, B.~T. 2003, \araa, 41, 241

\bibitem[{{Faherty} {et~al.}(2012){Faherty}, {Burgasser}, {Walter}, {Van der
  Bliek}, {Shara}, {Cruz}, {West}, {Vrba}, \&
  {Anglada-Escud{\'e}}}]{2012ApJ...752...56F}
{Faherty}, J.~K., {Burgasser}, A.~J., {Walter}, F.~M., {et~al.} 2012, \apj,
  752, 56

\bibitem[{{Filippazzo} {et~al.}(2015){Filippazzo}, {Rice}, {Faherty}, {Cruz},
  {Van Gordon}, \& {Looper}}]{2015ApJ...810..158F}
{Filippazzo}, J.~C., {Rice}, E.~L., {Faherty}, J., {et~al.} 2015, \apj, 810,
  158

\bibitem[{{Foreman-Mackey} {et~al.}(2013){Foreman-Mackey}, {Hogg}, {Lang}, \&
  {Goodman}}]{Foreman-Mackey2013emcee}
{Foreman-Mackey}, D., {Hogg}, D.~W., {Lang}, D., \& {Goodman}, J. 2013, \pasp,
  125, 306

\bibitem[{{Gagn{\'e}} {et~al.}(2015{\natexlab{a}}){Gagn{\'e}}, {Burgasser},
  {Faherty}, {Lafreni{\'e}re}, {Doyon}, {Filippazzo}, {Bowsher}, \&
  {Nicholls}}]{2015ApJ...808L..20G}
{Gagn{\'e}}, J., {Burgasser}, A.~J., {Faherty}, J.~K., {et~al.}
  2015{\natexlab{a}}, \apjl, 808, L20

\bibitem[{{Gagn{\'e}} {et~al.}(2014{\natexlab{a}}){Gagn{\'e}}, {Faherty},
  {Cruz}, {Lafreni{\`e}re}, {Doyon}, {Malo}, \&
  {Artigau}}]{2014ApJ...785L..14G}
{Gagn{\'e}}, J., {Faherty}, J.~K., {Cruz}, K., {et~al.} 2014{\natexlab{a}},
  \apjl, 785, L14

\bibitem[{{Gagn{\'e}} {et~al.}(2015{\natexlab{b}}){Gagn{\'e}}, {Faherty},
  {Cruz}, {Lafreni{\'e}re}, {Doyon}, {Malo}, {Burgasser}, {Naud}, {Artigau},
  {Bouchard}, {Gizis}, \& {Albert}}]{2015ApJS..219...33G}
{Gagn{\'e}}, J., {Faherty}, J.~K., {Cruz}, K.~L., {et~al.} 2015{\natexlab{b}},
  \apjs, 219, 33

\bibitem[{{Gagn{\'e}} {et~al.}(2014{\natexlab{b}}){Gagn{\'e}},
  {Lafreni{\`e}re}, {Doyon}, {Artigau}, {Malo}, {Robert}, \&
  {Nadeau}}]{2014ApJ...792L..17G}
{Gagn{\'e}}, J., {Lafreni{\`e}re}, D., {Doyon}, R., {et~al.}
  2014{\natexlab{b}}, \apjl, 792, L17

\bibitem[{{Gagn{\'e}} {et~al.}(2018){Gagn{\'e}}, {Mamajek}, {Malo}, {Riedel},
  {Rodriguez}, {Lafreni{\`e}re}, {Faherty}, {Roy-Loubier}, {Pueyo}, {Robin}, \&
  {Doyon}}]{2018arXiv180109051G}
{Gagn{\'e}}, J., {Mamajek}, E.~E., {Malo}, L., {et~al.} 2018, ArXiv e-prints
  [\eprint[arXiv]{1801.09051}]

\bibitem[{{Grether} \& {Lineweaver}(2006)}]{2006ApJ...640.1051G}
{Grether}, D. \& {Lineweaver}, C.~H. 2006, \apj, 640, 1051

\bibitem[{{Hagelberg} {et~al.}(2016){Hagelberg}, {S{\'e}gransan}, {Udry}, \&
  {Wildi}}]{2016MNRAS.455.2178H}
{Hagelberg}, J., {S{\'e}gransan}, D., {Udry}, S., \& {Wildi}, F. 2016, \mnras,
  455, 2178

\bibitem[{{Hinkley} {et~al.}(2013){Hinkley}, {Pueyo}, {Faherty}, {Oppenheimer},
  {Mamajek}, {Kraus}, {Rice}, {Ireland}, {David}, {Hillenbrand}, {Vasisht},
  {Cady}, {Brenner}, {Veicht}, {Nilsson}, {Zimmerman}, {Parry}, {Beichman},
  {Dekany}, {Roberts}, {Roberts}, {Baranec}, {Crepp}, {Burruss}, {Wallace},
  {King}, {Zhai}, {Lockhart}, {Shao}, {Soummer}, {Sivaramakrishnan}, \&
  {Wilson}}]{2013ApJ...779..153H}
{Hinkley}, S., {Pueyo}, L., {Faherty}, J.~K., {et~al.} 2013, \apj, 779, 153

\bibitem[{{H{\o}g} {et~al.}(2000){H{\o}g}, {Fabricius}, {Makarov}, {Urban},
  {Corbin}, {Wycoff}, {Bastian}, {Schwekendiek}, \&
  {Wicenec}}]{2000AA...355L..27H}
{H{\o}g}, E., {Fabricius}, C., {Makarov}, V.~V., {et~al.} 2000, \aap, 355, L27

\bibitem[{{Houk}(1993)}]{houk1993}
{Houk}, N. 1993, VizieR Online Data Catalog, 3051

\bibitem[{{Janson} {et~al.}(2012){Janson}, {Bonavita}, {Klahr}, \&
  {Lafreni{\`e}re}}]{2012ApJ...745....4J}
{Janson}, M., {Bonavita}, M., {Klahr}, H., \& {Lafreni{\`e}re}, D. 2012, \apj,
  745, 4

\bibitem[{{Janson} {et~al.}(2013){Janson}, {Lafreni{\`e}re}, {Jayawardhana},
  {Bonavita}, {Girard}, {Brandeker}, \& {Gizis}}]{2013ApJ...773..170J}
{Janson}, M., {Lafreni{\`e}re}, D., {Jayawardhana}, R., {et~al.} 2013, \apj,
  773, 170

\bibitem[{{Johnson} {et~al.}(2010){Johnson}, {Aller}, {Howard}, \&
  {Crepp}}]{2010PASP..122..905J}
{Johnson}, J.~A., {Aller}, K.~M., {Howard}, A.~W., \& {Crepp}, J.~R. 2010,
  \pasp, 122, 905

\bibitem[{{Jones} {et~al.}(2016){Jones}, {White}, {Quinn}, {Ireland},
  {Boyajian}, {Schaefer}, \& {Baines}}]{2016ApJ...822L...3J}
{Jones}, J., {White}, R.~J., {Quinn}, S., {et~al.} 2016, \apjl, 822, L3

\bibitem[{{Jones} {et~al.}(2014){Jones}, {Jenkins}, {Bluhm}, {Rojo}, \&
  {Melo}}]{2014A&A...566A.113J}
{Jones}, M.~I., {Jenkins}, J.~S., {Bluhm}, P., {Rojo}, P., \& {Melo}, C.~H.~F.
  2014, \aap, 566, A113

\bibitem[{{Kiraga}(2012)}]{kiraga2012}
{Kiraga}, M. 2012, \actaa, 62, 67

\bibitem[{{Kraus} {et~al.}(2008){Kraus}, {Ireland}, {Martinache}, \&
  {Lloyd}}]{kraus08}
{Kraus}, A.~L., {Ireland}, M.~J., {Martinache}, F., \& {Lloyd}, J.~P. 2008,
  \apj, 679, 762

\bibitem[{{Lachapelle} {et~al.}(2015){Lachapelle}, {Lafreni{\`e}re},
  {Gagn{\'e}}, {Jayawardhana}, {Janson}, {Helling}, \&
  {Witte}}]{2015ApJ...802...61L}
{Lachapelle}, F.-R., {Lafreni{\`e}re}, D., {Gagn{\'e}}, J., {et~al.} 2015,
  \apj, 802, 61

\bibitem[{{Lafreni{\`e}re} {et~al.}(2011){Lafreni{\`e}re}, {Jayawardhana},
  {Janson}, {Helling}, {Witte}, \& {Hauschildt}}]{2011ApJ...730...42L}
{Lafreni{\`e}re}, D., {Jayawardhana}, R., {Janson}, M., {et~al.} 2011, \apj,
  730, 42

\bibitem[{{Lafreni{\`e}re} {et~al.}(2008){Lafreni{\`e}re}, {Jayawardhana}, \&
  {van Kerkwijk}}]{2008ApJ...689L.153L}
{Lafreni{\`e}re}, D., {Jayawardhana}, R., \& {van Kerkwijk}, M.~H. 2008, \apjl,
  689, L153

\bibitem[{{Lagrange} {et~al.}(2010){Lagrange}, {Bonnefoy}, {Chauvin}, {Apai},
  {Ehrenreich}, {Boccaletti}, {Gratadour}, {Rouan}, {Mouillet}, {Lacour}, \&
  {Kasper}}]{2010Sci...329...57L}
{Lagrange}, A.-M., {Bonnefoy}, M., {Chauvin}, G., {et~al.} 2010, Science, 329,
  57

\bibitem[{{Lagrange} {et~al.}(2016){Lagrange}, {Langlois}, {Gratton}, {Maire},
  {Milli}, {Olofsson}, {Vigan}, {Bailey}, {Mesa}, {Chauvin}, {Boccaletti},
  {Galicher}, {Girard}, {Bonnefoy}, {Samland}, {Menard}, {Henning},
  {Kenworthy}, {Thalmann}, {Beust}, {Beuzit}, {Brandner}, {Buenzli},
  {Cheetham}, {Janson}, {le Coroller}, {Lannier}, {Mouillet}, {Peretti},
  {Perrot}, {Salter}, {Sissa}, {Wahhaj}, {Abe}, {Desidera}, {Feldt}, {Madec},
  {Perret}, {Petit}, {Rabou}, {Soenke}, \& {Weber}}]{2016A&A...586L...8L}
{Lagrange}, A.-M., {Langlois}, M., {Gratton}, R., {et~al.} 2016, \aap, 586, L8

\bibitem[{{Lannier} {et~al.}(2016){Lannier}, {Delorme}, {Lagrange}, {Borgniet},
  {Rameau}, {Schlieder}, {Gagn{\'e}}, {Bonavita}, {Malo}, {Chauvin},
  {Bonnefoy}, \& {Girard}}]{2016A&A...596A..83L}
{Lannier}, J., {Delorme}, P., {Lagrange}, A.~M., {et~al.} 2016, \aap, 596, A83

\bibitem[{{Leggett} {et~al.}(2000){Leggett}, {Allard}, {Dahn}, {Hauschildt},
  {Kerr}, \& {Rayner}}]{2000ApJ...535..965L}
{Leggett}, S.~K., {Allard}, F., {Dahn}, C., {et~al.} 2000, \apj, 535, 965

\bibitem[{{Liu} {et~al.}(2013){Liu}, {Dupuy}, \&
  {Allers}}]{2013AN....334...85L}
{Liu}, M.~C., {Dupuy}, T.~J., \& {Allers}, K.~N. 2013, Astronomische
  Nachrichten, 334, 85

\bibitem[{{Lodieu} {et~al.}(2008){Lodieu}, {Hambly}, {Jameson}, \&
  {Hodgkin}}]{2008MNRAS.383.1385L}
{Lodieu}, N., {Hambly}, N.~C., {Jameson}, R.~F., \& {Hodgkin}, S.~T. 2008,
  \mnras, 383, 1385

\bibitem[{{Lowrance} {et~al.}(2000){Lowrance}, {Schneider}, {Kirkpatrick},
  {Becklin}, {Weinberger}, {Zuckerman}, {Plait}, {Malmuth}, {Heap}, {Schultz},
  {Smith}, {Terrile}, \& {Hines}}]{2000ApJ...541..390L}
{Lowrance}, P.~J., {Schneider}, G., {Kirkpatrick}, J.~D., {et~al.} 2000, \apj,
  541, 390

\bibitem[{{Luhman} {et~al.}(2007){Luhman}, {Patten}, {Marengo}, {Schuster},
  {Hora}, {Ellis}, {Stauffer}, {Sonnett}, {Winston}, {Gutermuth}, {Megeath},
  {Backman}, {Henry}, {Werner}, \& {Fazio}}]{2007ApJ...654..570L}
{Luhman}, K.~L., {Patten}, B.~M., {Marengo}, M., {et~al.} 2007, \apj, 654, 570

\bibitem[{{Macintosh} {et~al.}(2008){Macintosh}, {Graham}, {Palmer}, {Doyon},
  {Dunn}, {Gavel}, {Larkin}, {Oppenheimer}, {Saddlemyer}, {Sivaramakrishnan},
  {Wallace}, {Bauman}, {Erickson}, {Marois}, {Poyneer}, \&
  {Soummer}}]{2008SPIE.7015E..18M}
{Macintosh}, B.~A., {Graham}, J.~R., {Palmer}, D.~W., {et~al.} 2008, in Society
  of Photo-Optical Instrumentation Engineers (SPIE) Conference Series, Vol.
  7015, Society of Photo-Optical Instrumentation Engineers (SPIE) Conference
  Series, 18

\bibitem[{{Madsen} {et~al.}(2002){Madsen}, {Dravins}, \&
  {Lindegren}}]{madsen2002}
{Madsen}, S., {Dravins}, D., \& {Lindegren}, L. 2002, \aap, 381, 446

\bibitem[{{Maire} {et~al.}(2016{\natexlab{a}}){Maire}, {Bonnefoy}, {Ginski},
  {Vigan}, {Messina}, {Mesa}, {Galicher}, {Gratton}, {Desidera}, {Kopytova},
  {Millward}, {Thalmann}, {Claudi}, {Ehrenreich}, {Zurlo}, {Chauvin},
  {Antichi}, {Baruffolo}, {Bazzon}, {Beuzit}, {Blanchard}, {Boccaletti}, {de
  Boer}, {Carle}, {Cascone}, {Costille}, {De Caprio}, {Delboulb{\'e}},
  {Dohlen}, {Dominik}, {Feldt}, {Fusco}, {Girard}, {Giro}, {Gisler}, {Gluck},
  {Gry}, {Henning}, {Hubin}, {Hugot}, {Jaquet}, {Kasper}, {Lagrange},
  {Langlois}, {Le Mignant}, {Llored}, {Madec}, {Martinez}, {Mawet}, {Milli},
  {M{\"o}ller-Nilsson}, {Mouillet}, {Moulin}, {Moutou}, {Orign{\'e}}, {Pavlov},
  {Petit}, {Pragt}, {Puget}, {Ramos}, {Rochat}, {Roelfsema}, {Salasnich},
  {Sauvage}, {Schmid}, {Turatto}, {Udry}, {Vakili}, {Wahhaj}, {Weber}, \&
  {Wildi}}]{2016A&A...587A..56M}
{Maire}, A.-L., {Bonnefoy}, M., {Ginski}, C., {et~al.} 2016{\natexlab{a}},
  \aap, 587, A56

\bibitem[{{Maire} {et~al.}(2016{\natexlab{b}}){Maire}, {Langlois}, {Dohlen},
  {Lagrange}, {Gratton}, {Chauvin}, {Desidera}, {Girard}, {Milli}, {Vigan},
  {Zins}, {Delorme}, {Beuzit}, {Claudi}, {Feldt}, {Mouillet}, {Puget},
  {Turatto}, \& {Wildi}}]{2016arXiv160906681M}
{Maire}, A.-L., {Langlois}, M., {Dohlen}, K., {et~al.} 2016{\natexlab{b}},
  \procspie

\bibitem[{{Marois} {et~al.}(2014){Marois}, {Correia}, {V{\'e}ran}, \&
  {Currie}}]{2014IAUS..299...48M}
{Marois}, C., {Correia}, C., {V{\'e}ran}, J.-P., \& {Currie}, T. 2014, in IAU
  Symposium, Vol. 299, IAU Symposium, ed. M.~{Booth}, B.~C. {Matthews}, \&
  J.~R. {Graham}, 48--49

\bibitem[{{Marois} {et~al.}(2006){Marois}, {Lafreni{\`e}re}, {Doyon},
  {Macintosh}, \& {Nadeau}}]{2006ApJ...641..556M}
{Marois}, C., {Lafreni{\`e}re}, D., {Doyon}, R., {Macintosh}, B., \& {Nadeau},
  D. 2006, \apj, 641, 556

\bibitem[{{Mawet} {et~al.}(2015){Mawet}, {David}, {Bottom}, {Hinkley},
  {Stapelfeldt}, {Padgett}, {Mennesson}, {Serabyn}, {Morales}, \&
  {Kuhn}}]{2015ApJ...811..103M}
{Mawet}, D., {David}, T., {Bottom}, M., {et~al.} 2015, \apj, 811, 103

\bibitem[{{McLean} {et~al.}(2003){McLean}, {McGovern}, {Burgasser},
  {Kirkpatrick}, {Prato}, \& {Kim}}]{2003ApJ...596..561M}
{McLean}, I.~S., {McGovern}, M.~R., {Burgasser}, A.~J., {et~al.} 2003, \apj,
  596, 561

\bibitem[{{Mesa} {et~al.}(2015){Mesa}, {Gratton}, {Zurlo}, {Vigan}, {Claudi},
  {Alberi}, {Antichi}, {Baruffolo}, {Beuzit}, {Boccaletti}, {Bonnefoy},
  {Costille}, {Desidera}, {Dohlen}, {Fantinel}, {Feldt}, {Fusco}, {Giro},
  {Henning}, {Kasper}, {Langlois}, {Maire}, {Martinez}, {Moeller-Nilsson},
  {Mouillet}, {Moutou}, {Pavlov}, {Puget}, {Salasnich}, {Sauvage}, {Sissa},
  {Turatto}, {Udry}, {Vakili}, {Waters}, \& {Wildi}}]{2015A&A...576A.121M}
{Mesa}, D., {Gratton}, R., {Zurlo}, A., {et~al.} 2015, \aap, 576, A121

\bibitem[{{Metchev} \& {Hillenbrand}(2009)}]{2009ApJS..181...62M}
{Metchev}, S.~A. \& {Hillenbrand}, L.~A. 2009, \apjs, 181, 62

\bibitem[{{Nielsen} {et~al.}(2012){Nielsen}, {Liu}, {Wahhaj}, {Biller},
  {Hayward}, {Boss}, {Bowler}, {Kraus}, {Shkolnik}, {Tecza}, {Chun}, {Clarke},
  {Close}, {Ftaclas}, {Hartung}, {Males}, {Reid}, {Skemer}, {Alencar},
  {Burrows}, {de Gouveia Dal Pino}, {Gregorio-Hetem}, {Kuchner}, {Thatte}, \&
  {Toomey}}]{2012ApJ...750...53N}
{Nielsen}, E.~L., {Liu}, M.~C., {Wahhaj}, Z., {et~al.} 2012, \apj, 750, 53

\bibitem[{{Olofsson} {et~al.}(2016){Olofsson}, {Samland}, {Avenhaus},
  {Caceres}, {Henning}, {Mo{\'o}r}, {Milli}, {Canovas}, {Quanz}, {Schreiber},
  {Augereau}, {Bayo}, {Bazzon}, {Beuzit}, {Boccaletti}, {Buenzli}, {Casassus},
  {Chauvin}, {Dominik}, {Desidera}, {Feldt}, {Gratton}, {Janson}, {Lagrange},
  {Langlois}, {Lannier}, {Maire}, {Mesa}, {Pinte}, {Rouan}, {Salter},
  {Thalmann}, \& {Vigan}}]{2016A&A...591A.108O}
{Olofsson}, J., {Samland}, M., {Avenhaus}, H., {et~al.} 2016, \aap, 591, A108

\bibitem[{{Patience} {et~al.}(2010){Patience}, {King}, {de Rosa}, \&
  {Marois}}]{2010A&A...517A..76P}
{Patience}, J., {King}, R.~R., {de Rosa}, R.~J., \& {Marois}, C. 2010, \aap,
  517, A76

\bibitem[{{Pavlov} {et~al.}(2008){Pavlov}, {M{\"o}ller-Nilsson}, {Feldt},
  {Henning}, {Beuzit}, \& {Mouillet}}]{2008SPIE.7019E..39P}
{Pavlov}, A., {M{\"o}ller-Nilsson}, O., {Feldt}, M., {et~al.} 2008, in
  \procspie, Vol. 7019, Advanced Software and Control for Astronomy II, 701939

\bibitem[{{Pecaut} \& {Mamajek}(2013)}]{pecaut2013}
{Pecaut}, M.~J. \& {Mamajek}, E.~E. 2013, \apjs, 208, 9

\bibitem[{{Pecaut} \& {Mamajek}(2016)}]{pecaut2016}
{Pecaut}, M.~J. \& {Mamajek}, E.~E. 2016, \mnras, 461, 794

\bibitem[{{Rayner} {et~al.}(2009){Rayner}, {Cushing}, \&
  {Vacca}}]{2009ApJS..185..289R}
{Rayner}, J.~T., {Cushing}, M.~C., \& {Vacca}, W.~D. 2009, \apjs, 185, 289

\bibitem[{{Rizzuto} {et~al.}(2011){Rizzuto}, {Ireland}, \&
  {Robertson}}]{rizzuto2011}
{Rizzuto}, A.~C., {Ireland}, M.~J., \& {Robertson}, J.~G. 2011, \mnras, 416,
  3108

\bibitem[{{Robert} {et~al.}(2016){Robert}, {Gagn{\'e}}, {Artigau},
  {Lafreni{\`e}re}, {Nadeau}, {Doyon}, {Malo}, {Albert}, {Simard}, {Bardalez
  Gagliuffi}, \& {Burgasser}}]{2016ApJ...830..144R}
{Robert}, J., {Gagn{\'e}}, J., {Artigau}, {\'E}., {et~al.} 2016, \apj, 830, 144

\bibitem[{Skrutskie {et~al.}(2006)Skrutskie, Cutri, Stiening, Weinberg,
  Schneider, Carpenter, Beichman, Capps, Chester, Elias,
  {et~al.}}]{skrutskie2006two}
Skrutskie, M., Cutri, R., Stiening, R., {et~al.} 2006, \aj, 131, 1163

\bibitem[{{Slawson} {et~al.}(1992){Slawson}, {Hill}, \&
  {Landstreet}}]{slawson1992}
{Slawson}, R.~W., {Hill}, R.~J., \& {Landstreet}, J.~D. 1992, \apjs, 82, 117

\bibitem[{{Soummer} {et~al.}(2012){Soummer}, {Pueyo}, \&
  {Larkin}}]{2012ApJ...755L..28S}
{Soummer}, R., {Pueyo}, L., \& {Larkin}, J. 2012, \apjl, 755, L28

\bibitem[{{Toomre}(1964)}]{1964ApJ...139.1217T}
{Toomre}, A. 1964, \apj, 139, 1217

\bibitem[{{Torres} {et~al.}(2006){Torres}, {Quast}, {da Silva}, {de La Reza},
  {Melo}, \& {Sterzik}}]{sacy}
{Torres}, C.~A.~O., {Quast}, G.~R., {da Silva}, L., {et~al.} 2006, \aap, 460,
  695

\bibitem[{{van Leeuwen}(2007)}]{vl2007}
{van Leeuwen}, F. 2007, \aap, 474, 653

\bibitem[{{Vigan}(2016)}]{2016ascl.soft03001V}
{Vigan}, A. 2016, {SILSS: SPHERE/IRDIS Long-Slit Spectroscopy pipeline},
  Astrophysics Source Code Library

\bibitem[{{Vigan} {et~al.}(2008){Vigan}, {Langlois}, {Moutou}, \&
  {Dohlen}}]{2008A&A...489.1345V}
{Vigan}, A., {Langlois}, M., {Moutou}, C., \& {Dohlen}, K. 2008, \aap, 489,
  1345

\bibitem[{{Vigan} {et~al.}(2010){Vigan}, {Moutou}, {Langlois}, {Allard},
  {Boccaletti}, {Carbillet}, {Mouillet}, \& {Smith}}]{2010MNRAS.407...71V}
{Vigan}, A., {Moutou}, C., {Langlois}, M., {et~al.} 2010, \mnras, 407, 71

\bibitem[{{Vigan} {et~al.}(2016){Vigan}, {N'Diaye}, {Dohlen}, {Beuzit},
  {Costille}, {Caillat}, {Baruffolo}, {Blanchard}, {Carle}, {Ferrari}, {Fusco},
  {Gluck}, {Hugot}, {Jaquet}, {Langlois}, {Le Mignant}, {Llored}, {Madec},
  {Mouillet}, {Orign{\'e}}, {Puget}, {Salasnich}, \&
  {Sauvage}}]{2016SPIE.9912E..26V}
{Vigan}, A., {N'Diaye}, M., {Dohlen}, K., {et~al.} 2016, in \procspie, Vol.
  9912, Advances in Optical and Mechanical Technologies for Telescopes and
  Instrumentation II, 991226

\bibitem[{{Vigan} {et~al.}(2012){Vigan}, {Patience}, {Marois}, {Bonavita}, {De
  Rosa}, {Macintosh}, {Song}, {Doyon}, {Zuckerman}, {Lafreni{\`e}re}, \&
  {Barman}}]{2012A&A...544A...9V}
{Vigan}, A., {Patience}, J., {Marois}, C., {et~al.} 2012, \aap, 544, A9

\bibitem[{{Wright} {et~al.}(2010){Wright}, {Eisenhardt}, {Mainzer}, {Ressler},
  {Cutri}, {Jarrett}, {Kirkpatrick}, {Padgett}, {McMillan}, {Skrutskie},
  {Stanford}, {Cohen}, {Walker}, {Mather}, {Leisawitz}, {Gautier}, {McLean},
  {Benford}, {Lonsdale}, {Blain}, {Mendez}, {Irace}, {Duval}, {Liu}, {Royer},
  {Heinrichsen}, {Howard}, {Shannon}, {Kendall}, {Walsh}, {Larsen}, {Cardon},
  {Schick}, {Schwalm}, {Abid}, {Fabinsky}, {Naes}, \&
  {Tsai}}]{2010AJ....140.1868W}
{Wright}, E.~L., {Eisenhardt}, P.~R.~M., {Mainzer}, A.~K., {et~al.} 2010, \aj,
  140, 1868

\bibitem[{{Zurlo} {et~al.}(2016){Zurlo}, {Vigan}, {Galicher}, {Maire}, {Mesa},
  {Gratton}, {Chauvin}, {Kasper}, {Moutou}, {Bonnefoy}, {Desidera}, {Abe},
  {Apai}, {Baruffolo}, {Baudoz}, {Baudrand}, {Beuzit}, {Blancard},
  {Boccaletti}, {Cantalloube}, {Carle}, {Cascone}, {Charton}, {Claudi},
  {Costille}, {de Caprio}, {Dohlen}, {Dominik}, {Fantinel}, {Feautrier},
  {Feldt}, {Fusco}, {Gigan}, {Girard}, {Gisler}, {Gluck}, {Gry}, {Henning},
  {Hugot}, {Janson}, {Jaquet}, {Lagrange}, {Langlois}, {Llored}, {Madec},
  {Magnard}, {Martinez}, {Maurel}, {Mawet}, {Meyer}, {Milli},
  {Moeller-Nilsson}, {Mouillet}, {Orign{\'e}}, {Pavlov}, {Petit}, {Puget},
  {Quanz}, {Rabou}, {Ramos}, {Rousset}, {Roux}, {Salasnich}, {Salter},
  {Sauvage}, {Schmid}, {Soenke}, {Stadler}, {Suarez}, {Turatto}, {Udry},
  {Vakili}, {Wahhaj}, {Wildi}, \& {Antichi}}]{2016A&A...587A..57Z}
{Zurlo}, A., {Vigan}, A., {Galicher}, R., {et~al.} 2016, \aap, 587, A57

\end{thebibliography}

\begin{appendix}

\section{Upper limits on the dust mass\label{appendix:dust_mass}}

\citet{2012ApJ...756..133C} reported a non-detection of a mid-IR excess around HIP\,64892, with a \textit{Spitzer}/MIPS upper limit at 70\,$\mu$m of $15.8$\,mJy. Using a similar approach to that of \cite{2017A&A...605L...9C} for HIP 65426, we can convert this value to an upper limit on the dust mass around HIP 64892. Similar to HIP 65426, this upper limit does not reach the photospheric flux ($\sim1.4$\,mJy).

\begin{figure}
\includegraphics[width=\columnwidth]{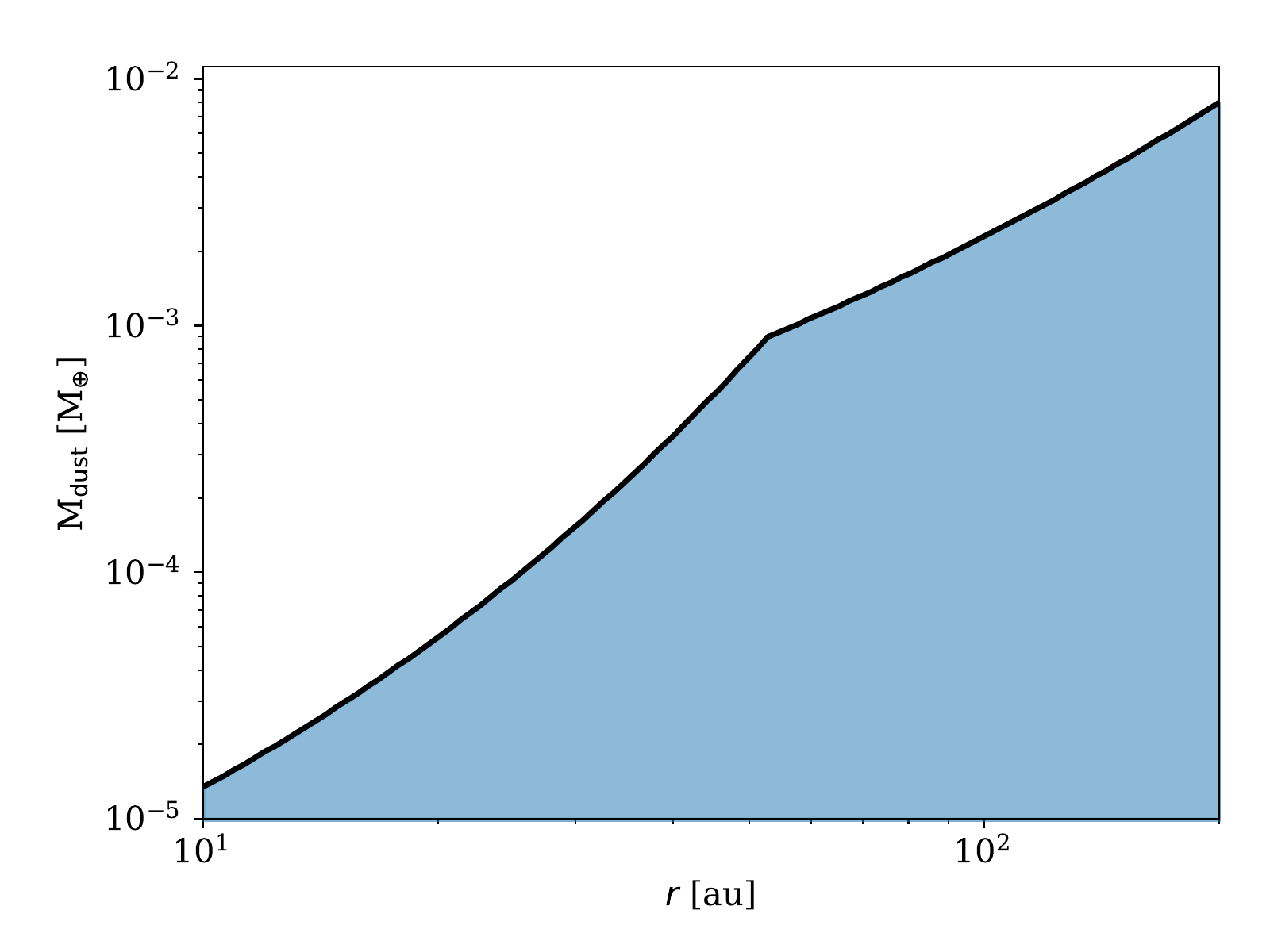}
\caption{The blue-shaded area shows the region in the $r$-$M_{\mathrm{dust}}$ plane, in which a debris disk could be present and remain compatible with the mid- and far-IR observations.}
\label{fig:mdust}
\end{figure}

We gather the optical to mid-IR photometry of the star using VOSA\footnote{http://svo2.cab.inta-csic.es/theory/vosa/} \citep{2008A&A...492..277B}. For the given stellar luminosity and mass ($33$\,$L_{\odot}$ and $2.35$\,$M_{\odot}$, respectively) we estimate the size of dust grains that would still be on bound orbits around the star. We use the optical constant of astro-silicates \citep{2003ARA&A..41..241D}, and we compute the radiation pressure to gravitational forces $\beta$ ratio as in \citet{1979Icar...40....1B}. We find that for this composition, grains larger than $s_{\mathrm{blow}} \sim 4.8$\,$\mu$m should remain on bound orbits around the star. To estimate the possible configurations for a debris disk to remain compatible with the mid- and far-IR observations, we compute a series of disk models \citep[similar to][]{2016A&A...591A.108O}. We consider a grain size distribution of the form d$n(s) \propto s^{-3.5}$d$s$, between $s_{\mathrm{min}} = s_{\mathrm{blow}}$ and $s_{\mathrm{max}} = 1$\,mm. We sample $100$ $r_{\mathrm{i}}$ values for the radial distance of the belt between $10$ and $200$\,au. For each $r_{\mathrm{i}}$, we consider a disk model between $0.9 \times r_{\mathrm{i}} \leq r_{\mathrm{i}} \leq 1.1 \times r_{\mathrm{i}}$. We then slowly increase the mass of the disk until the thermal emission (plus the stellar contribution) is larger than either the \textit{WISE}/W4 $22$\,$\mu$m point or the \textit{Spitzer}/MIPS 70\,$\mu$m point. We therefore delimit a region in the $r$-$M_{\mathrm{dust}}$ plane where debris disks could exist and remain undetected with the current observations (see Fig.\,\ref{fig:mdust}). 
Overall, with our assumptions on the radial extent of the debris disk, we find the dust mass must be less than $\sim 2 \times 10^{-3}$\,M$_{\oplus}$ at about $100$\,au from the star.
When compared to the well-known $\beta$ Pictoris debris disk, we find that any potential belts around HIP 64892 must be substantially less massive. The total dust mass of the $\beta$ Pictoris debris disk was measured at $\sim8\times 10^{-2}$\,M$_{\oplus}$ \citep{2014Sci...343.1490D} between 50-120AU, implying a value of $\sim3\times10^{-2}$\,M$_{\oplus}$ for an annulus at 100AU that can be compared with our simulation (assuming constant density).

\section{Disk Instability models for HIP 64892B \label{app:gi_models}}
\begin{figure}
\includegraphics[width=0.98\columnwidth]{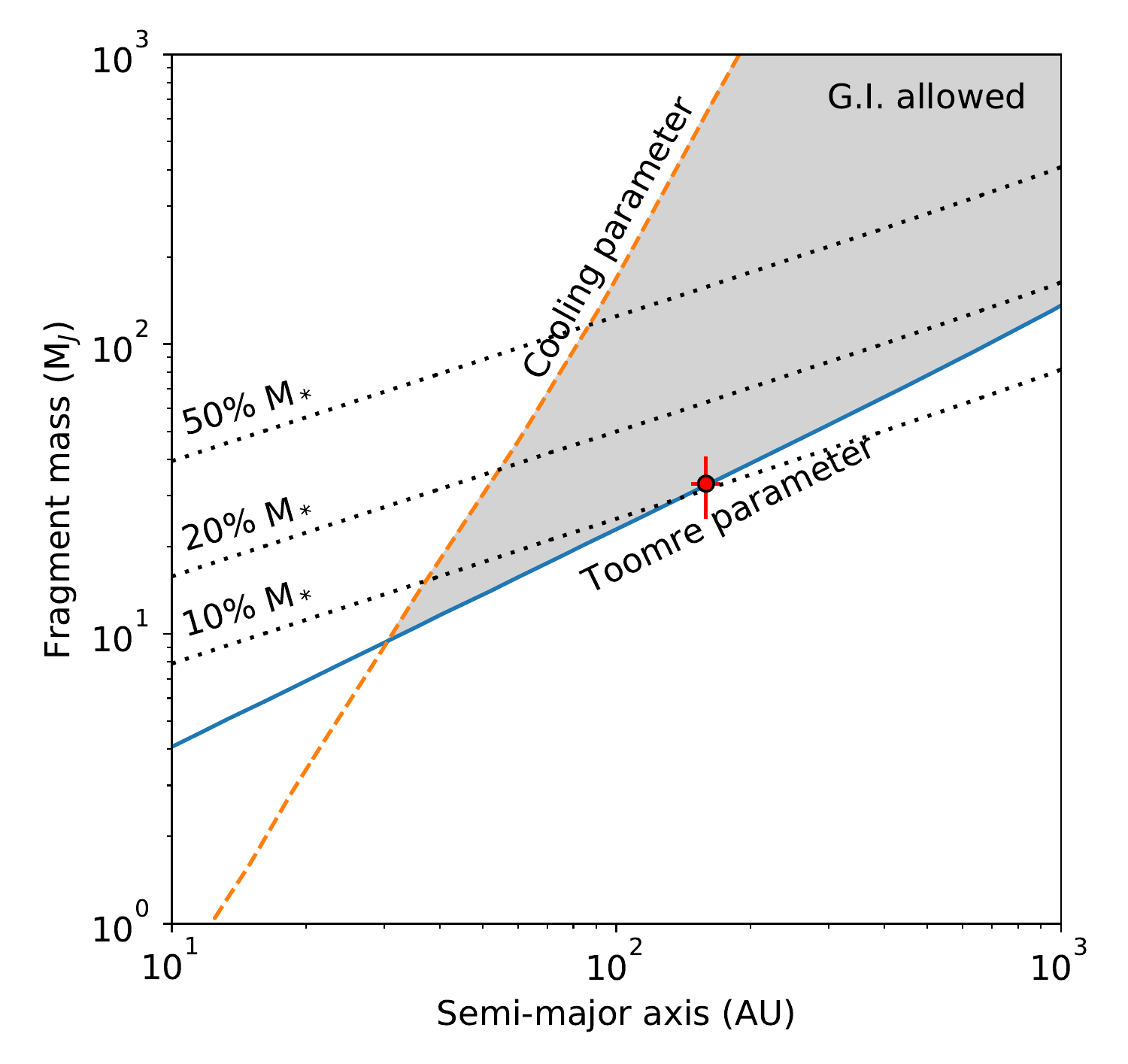}
\caption{Predictions for the fragment masses compatible with production via gravitational instability for HIP 64892. Fragments with masses above the orange dashed line cannot cool efficiently enough, while those below the blue solid line do not satisfy the Toomre criterion. The mass and projected separation of HIP 64892B are marked. We find that HIP 64892B is compatible with in-situ formation via gravitational instability, although further constraints on the semi-major axis are needed to confirm this idea. The initial disk masses required to form fragments of a given size are shown with black dotted lines.}\label{fig:gi_model}
\end{figure}

To investigate the possibility that HIP 64892B formed via disk instability we estimate the range of fragment masses that could be produced as a function of semi-major axis, using a set of fragmentation criteria as recently confirmed in local high resolution 3D simulations \citep{2017ApJ...848...40B}. These models are described in detail in \citet{2012ApJ...745....4J} and \citet{ 2014A&A...562A.111B}. Briefly, fragments must satisfy the Toomre criterion for self-gravitating clumps \citep{1964ApJ...139.1217T} and be able to cool faster than the local Keplerian timescale. These models require the star's initial luminosity and metallicity. The former was estimated for HIP 64892 from the isochrones of \citet{bressan2012}, while for the latter we assumed a solar metallicity. The result is plotted in Fig. \ref{fig:gi_model}. 
We also compared the primordial disk mass required to support fragments of a given mass, for disks with 10\%, 20\% and 50\% of the mass of the host star.

We found that if the currently observed projected separation of HIP 64892B is close to its semi-major axis, then its predicted mass and location are close to the Toomre limit. This opens the possibility that the companion may have formed in-situ via gravitational instability.

The same models were applied by \citet{2014A&A...562A.111B} to the companions $\kappa$ And B, HR 7329B, HD 1160B, and HIP 78530B. These objects are all young substellar companions to 2-2.5\,M$_\odot$ stars. HIP 64892B falls in a similar region of the diagram to $\kappa$ And B, HR 7329B and HD 1160B. These 4 objects are compatible with in-situ formation via disk instability, and require only modest disk masses of 10-20\% of the mass of the primary star to form. HIP 78530B is a clear outlier at a much larger separation that puts it below the Toomre limit, indicating that its formation proceeded through a different pathway, or that it underwent significant outward orbital migration since its formation epoch.

\end{appendix}
\end{document}